\documentclass[12pt,preprint]{aastex}

\newcommand{\etal}{et\,al.}
\newcommand{\halpha}{H$\alpha$}
\newcommand{\lsim}{\raise0.3ex\hbox{$<$}\kern-0.75em{\lower0.65ex\hbox{$\sim$}}}

\newcommand{\msun}{M$_{\odot}$}
\newcommand{\HII}{H~{\sc ii}}
\newcommand{\HI}{H~{\sc I}}
\newcommand{\kms}{km\,s$^{-1}$}
\newcommand{\pom}{\,$\pm$\,}

\newcommand{\gsim}{\raise0.3ex\hbox{$>$}\kern-0.75em{\lower0.65ex\hbox{$\sim$}}}
\begin{document}
\slugcomment{The Astrophysical Journal, in press}

\title{The M81 Group Dwarf Irregular Galaxy DDO 165. II.  Connecting
Recent Star Formation with ISM Structures and
Kinematics\footnote{Based on observations made with the NASA/ESA
Hubble Space Telescope, obtained from the Data Archive at the Space
Telescope Science Institute, which is operated by the Association of
Universities for Research in Astronomy, Inc., under NASA contract NAS
5-26555.}}
 



\author{John M. Cannon, Hans P. Most}
\affil{Department of Physics \& Astronomy, Macalester College, 1600 Grand Avenue, Saint Paul, MN 55105}
\email{jcannon@macalester.edu; hmost@macalester.edu}

\author{Evan D. Skillman, Daniel R. Weisz\footnote{Current address: Department of Astronomy, Box 351580, University of Washington, Seattle, WA 98195, USA}}
\affil{Astronomy Department, University of Minnesota, Minneapolis, MN 55455}
\email{skillman@astro.umn.edu, dweisz@astro.umn.edu}

\author{David Cook}
\affil{Department of Physics and Astronomy, University of Wyoming, Laramie, WY 82071, USA}
\email{dcook12@uwyo.edu}

\author{Andrew E. Dolphin}
\affil{Raytheon Company, 1151 East Hermans Road, Tucson, AZ 85756}
\email{adolphin@raytheon.com}

\author{Robert C. Kennicutt, Jr.}
\affil{Institute of Astronomy, University of Cambridge, Madingley Road,
Cambridge, CB3 0HA, United Kingdom}
\email{robk@ast.cam.ac.uk}

\author{Janice Lee}
\affil{Observatories of the Carnegie Institution of Washington, 813 Santa Barbara Street, Pasadena, CA 91101}
\email{jlee@obs.carnegiescience.edu}

\author{Anil Seth\footnote{OIR Fellow}}
\affil{Harvard-Smithsonian Center for Astrophysics, 60 Garden Street Cambridge, MA 02138}
\email{aseth@cfa.harvard.edu}

\author{Fabian Walter}
\affil{Max-Planck-Institut f{\"u}r Astronomie, K{\"o}nigstuhl 17, D-69117, Heidelberg, Germany}
\email{walter@mpia.de}

\author{Steven R. Warren}
\affil{Astronomy Department, University of Minnesota, Minneapolis, MN 55455}
\email{warren@astro.umn.edu}

\begin{abstract}

We compare the stellar populations and complex neutral gas dynamics of
the M81 group dIrr galaxy DDO\,165 using data from the {\it HST} and
the {\it VLA}.  Paper I identified two kinematically distinct
\HI\ components, multiple localized high velocity gas features, and
eight \HI\ holes and shells (the largest of which spans
$\sim$2.2\,$\times$\,1.1 kpc).  Using the spatial and temporal
information from the stellar populations in DDO\,165, we compare the
patterns of star formation over the past 500 Myr with the
\HI\ dynamics.  We extract localized star formation histories within 6
of the 8 \HI\ holes identified in Paper I, as well as 23 other regions
that sample a range of stellar densities and neutral gas properties.
From population synthesis modeling, we derive the energy outputs
(from stellar winds and supernovae) of the stellar populations within
these regions over the last 100 Myr, and compare with refined
estimates of the energies required to create the \HI\ holes.  In all
cases, we find that ``feedback'' is energetically capable of creating
the observed structures in the ISM.  Numerous regions with significant
energy inputs from feedback lack coherent \HI\ structures but show
prominent localized high velocity gas features; this feedback
signature is a natural product of temporally and spatially distributed
star formation.  In DDO\,165, the extended period of heightened star
formation activity (lasting more than 1 Gyr) is energetically capable
of creating the observed holes and high velocity gas features in the
neutral ISM.

\end{abstract}						


\section{Introduction}
\label{S1}

As stars evolve, the energies from stellar winds and supernova (SN)
explosions (hereafter collectively referred to as ``feedback'') are
released into the surrounding interstellar material.  The physics of
this interaction is complex, with dependencies on (at least) the local
neutral and molecular gas density and porosity, the local and total
gravitational potential depth and shape, the intensity and duration of
massive star formation (SF), and the metal content of the parent
cloud.  In a simplistic scenario, a single massive star forms and
evolves, unleashing a characteristic energy of $\sim$10$^{51}$ erg per
SN event \citep{burrows00}.  A fraction of this energy is converted
into bulk motion of the surrounding material.  This energy can be
compared with the amount of energy required to move that same gas mass
a certain distance within a given gravitational potential.  If the
feedback energy exceeds the requisite formation energy, then stellar
evolution is capable of creating structures in the interstellar medium
(ISM).

Scaling this simple single-star scenario up to bursts of SF, the
conditions for the creation of macroscopic ISM features could be
idealized in the case of an intense and rapid SF event
\citep{maclow99}.  This would result in the simultaneous and coherent
injection of feedback energy from multiple stars.  One would expect to
find coherent ISM structures, such as \HI\ holes and shells, with
remnant stellar clusters within; those cluster ages would in principle
be similar to the age of the ISM feature that surrounds it.

Recent results have challenged this simplistic model.  In systems
littered with coherent \HI\ structures (e.g., Holmberg\,II; see
discussion in {Weisz \etal\ 2009b}\nocite{weisz09b}), there are few
instances of high surface brightness remnant clusters (see also {Rhode
\etal\ 1999}\nocite{rhode99}).  However, deep {\it Hubble
Space Telescope} ({\it HST}) imaging reveals young stellar
populations within all \HI\ features; energy estimates based on
spatially resolved star formation history (SFH) analyses show that the
energy requirements for the creation of those ISM features is
easily met by stellar winds and supernovae.  Importantly, this energy
is derived from SF spanning hundreds of Myr and encompassing multiple
episodes of SF within a given region.  A new scenario supported by
recent work is one where the present-day morphology and dynamics of a
given galaxy are the results of the cumulative SF over at least the
past few hundred Myr.

\begin{figure}
\epsscale{1}
\plotone{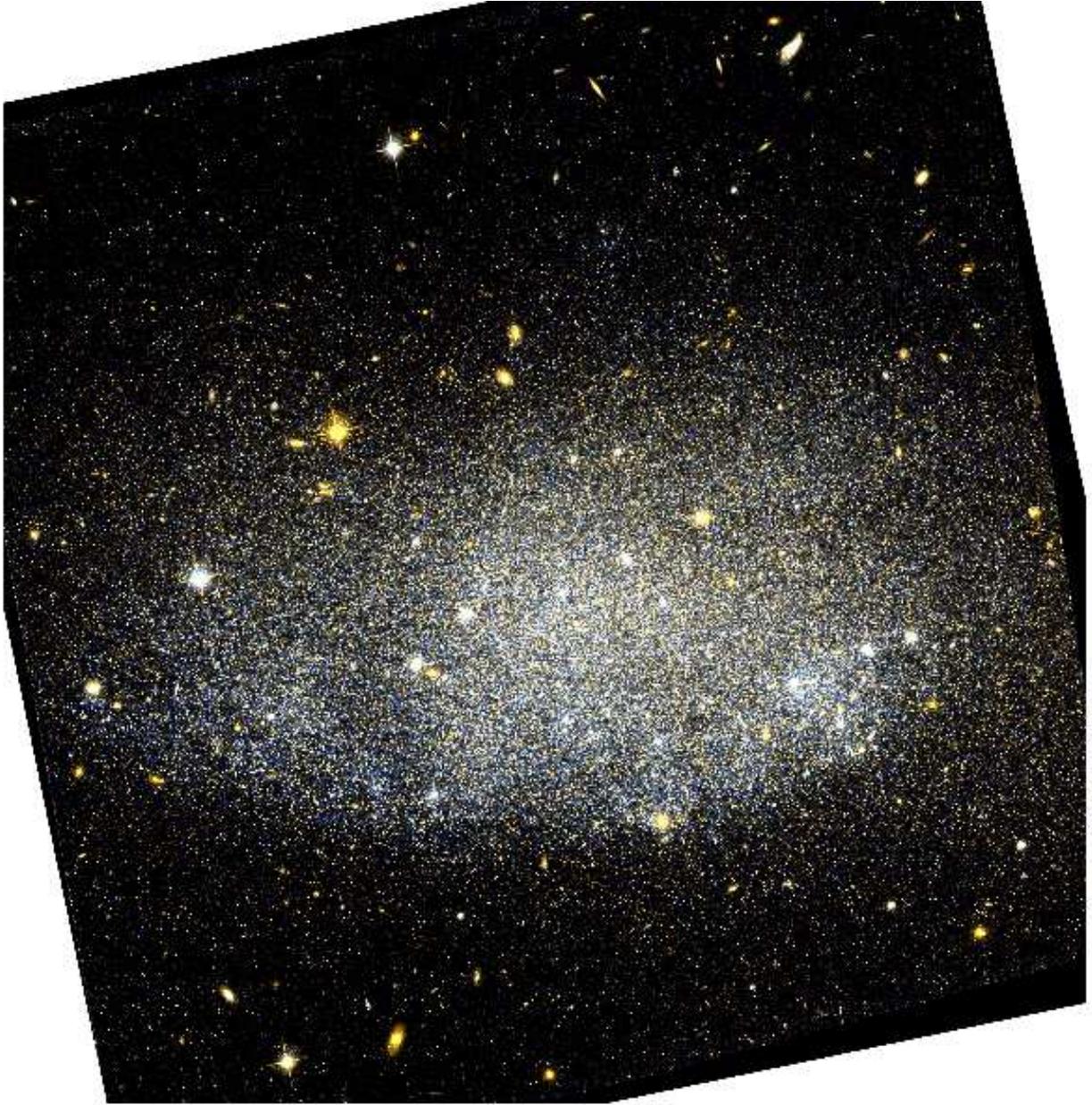}
\epsscale{1.0}
\caption{An {\it HST}/ACS color image of DDO\,165, created using F555W as
blue and F814W as red; north is up and east is to the left.  Note the
significant blue stellar population throughout the disk.}
\label{figcap1}
\end{figure}

The properties of the stellar populations in DDO\,165 make it a
particularly important test case for this model of the SF-ISM
interaction.  As the color {\it HST} image in Figure~\ref{figcap1}
shows, DDO\,165 contains rich populations of luminous stars, both blue
and red.  The global color magnitude diagram (CMD; see
Figure~\ref{figcap2}) derived from {\it HST} data shows that the
system has been in a burst phase (defined as the SF rate at recent
times exceeding the average past rate by more than factor of two) for
more than 1 Gyr \citep[see details in][]{mcquinn10}.  This easily
exceeds the dynamical timescale (710\,\pom\,40 Myr) and is among the
longest starburst episodes known to date in any dwarf galaxy.  The
recent (up to 500 Myr) SFH shows significant variations on timescales
$\lsim$50 Myr, including a precipitous drop in SF activity within the
past 25 Myr (see detailed discussion below). Further,
\citet{bastian11} find that DDO\,165 has a long characteristic
timescale for structure evolution in its stellar populations (at least
350 Myr; see further discussion below); once formed, stellar
associations remain coherent for a large fraction of a dynamical
timescale.

The neutral ISM dynamics of DDO\,165 (see Cannon \etal\ 2011;
hereafter Paper I) are equally intriguing in the context of the
feedback model.  The \HI\ morphology is dominated by a giant \HI\ hole
(diameter $\sim$2.2\,$\times$\,1.1 kpc) that is slightly offset from
the center of the stellar distribution.  Seven other \HI\ holes are
cataloged in Paper I; some of these have measurable expansion
velocities, while others are static at the present time.  The \HI\ gas
is concentrated in two kinematically distinct features: a southern,
main component and a northern component that contains $\sim$15\% of
the total \HI\ mass.  Based on the \HI\ data alone, we can exclude
many potential origins for this kinematic discontinuity (e.g.,
coherent solid-body rotation; superposed or counter-rotating disks;
turbulence), but both infall (i.e., interaction) and outflow (i.e.,
``blowout'') models remain viable.  The southern \HI\ component
harbors multiple localized high velocity features, each of which spans
up to $\sim$60 \kms\ in velocity space but is no larger than the
\HI\ beam width ($\sim$430 pc).

Taken together, these peculiar properties of DDO\,165 make it a
promising laboratory for examining the effects of feedback on the
surrounding interstellar material.  To that end, the present work
investigates the stellar populations throughout DDO\,165, exploiting
the spatially resolved {\it HST} data to study those populations
associated with each of the features identified in the \HI\ analysis
of Paper I.  In a broad sense, we seek to understand if spatially and
temporally extended SF is capable of creating giant \HI\ holes, high
velocity gas features, and kinematically distinct gas components.

\section{Observations and Data Reduction}
\label{S2}

DDO\,165 was observed with the {\it Very Large Array\footnote{The
    National Radio Astronomy Observatory (NRAO) is a facility of the
    National Science Foundation operated under cooperative agreement
    by Associated Universities, Inc.}} ({\it VLA}) in the
\HI\ spectral line during three observing sessions between November,
2006 and November, 2007 for program AC842 (P.I. Cannon).  The
treatment of these data, including reductions, handling and analyses,
are described fully in Paper I.

DDO\,165 was observed with the {\it HST} on March 17, 2006, as part of
observing program GO-10605 (P.I. Skillman).  The treatment and
reduction of these images are discussed in detail in \citet{weisz08}.
Over 144,000 stars are measured using the DOLPHOT package
\citep{dolphin00}; the 50\% completeness levels are m$_{\rm I}=$28.0
(M$_{\rm I}=-$0.3) and m$_{\rm V}=$27.2 (M$_{\rm V}=-$1.1).

\section{The Stellar Populations of DDO\,165}
\label{S3}

\subsection{Global Color-Magnitude Diagram}
\label{S3.1}

Figure~\ref{figcap1} shows a color representation of the {\it HST}/ACS
field of view of DDO\,165.  All of the high surface brightness
stellar populations are covered by this single pointing.  Note the
pronounced truncation of the young stellar population in the southern
region of the system (see also further discussion in \S~\ref{S4.1}).
The brightest objects in the field are Galactic foreground stars;
other compact high surface brightness objects are stellar clusters
within DDO\,165 (see further discussion in \S~\ref{S3.3.3} below) or
background galaxies.

\begin{figure}
\plotone{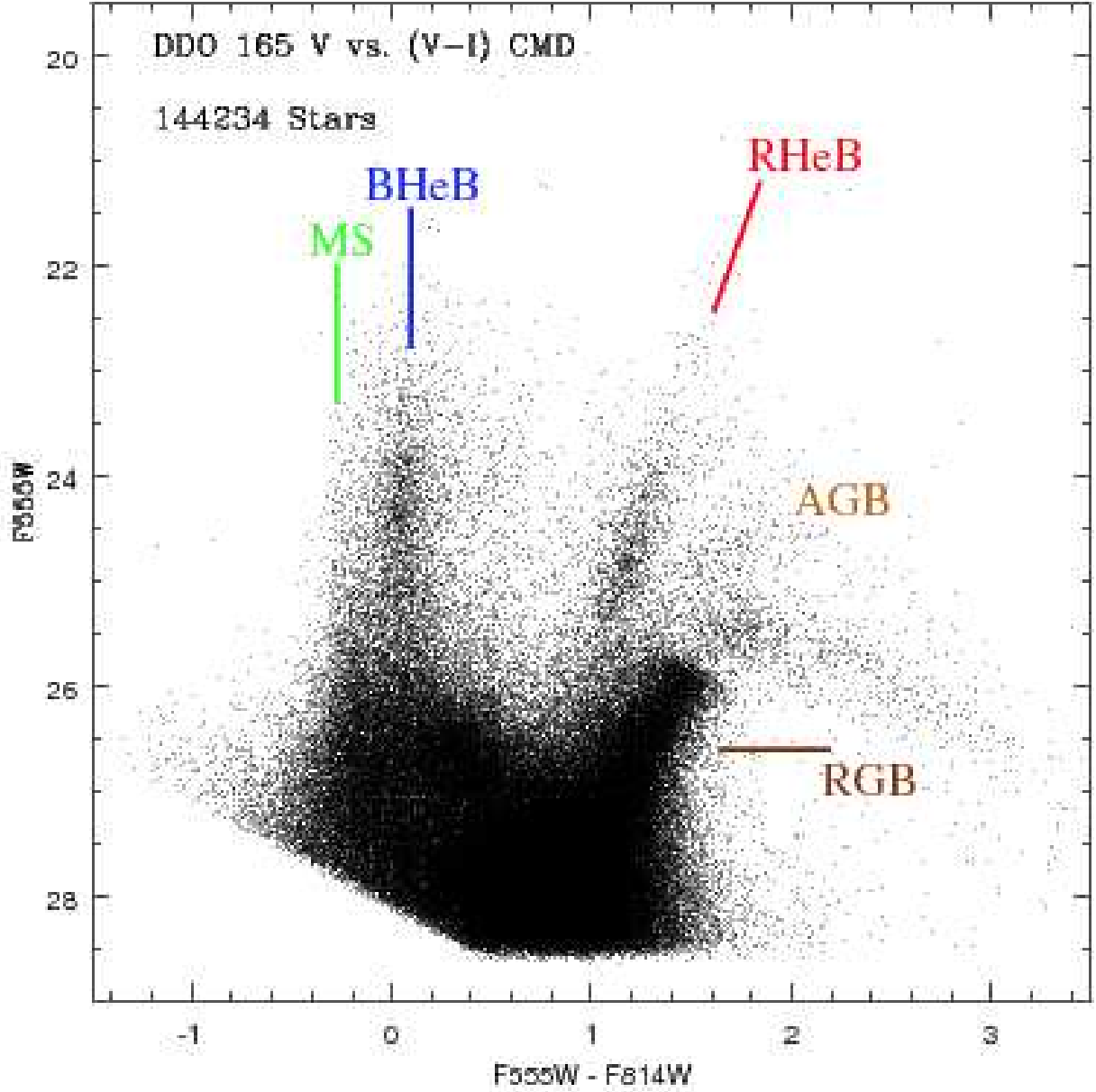}
\epsscale{1.0}
\caption{V vs. (V$-$I) CMD of all stars in DDO\,165.  The main phases
  of stellar evolution are labeled to guide the eye.  The BHeB
  sequence is well-populated throughout DDO\,165; the comparatively
  weak MS emphasizes the ``post-starburst'' nature of the system.}
\label{figcap2}
\end{figure}

The V vs.\ (V$-$I) CMD of all 144,234 stars within the {\it HST}/ACS
field of view is shown in Figure~\ref{figcap2}.  Five prominent
stellar evolution phases are indicated: the main sequence (MS), the
blue helium burning sequence (BHeB), the red helium burning sequence
(RHeB), the asymptotic giant branch (AGB), and the red giant branch
(RGB).  Note the strengths of the BHeB and RHeB sequences compared to
the MS.  The BHeB is $\sim$2 magnitudes brighter than the MS at a
given stellar age; thus, the prominence of the helium burning sequences
suggests stronger SF in the recent past than at the current time.

In this work, we focus on the BHeB stars because of their unique
relationship between age and absolute magnitude.  As discussed in
detail in \citet{dohmpalmer97} and multiple works since then, the ages
of BHeB stars increase as the absolute magnitude increases.  Further,
the age gradient becomes shallower for increasingly faint BHeB stars,
resulting in degraded temporal resolution at larger look back times;
for example, 10 Myr resolution is available back to $\sim$100 Myr, but
$\sim$50 Myr time bins are required when going back to 500 Myr.  The
BHeB merges with the red clump at ages $\gsim$600 Gyr; at larger
lookback times, other regions of the CMD must be used.  In the present
work we concentrate on the most recent 500 Myr, during which the SFH
is quite robust. The observed CMDs are discussed here and in
\S~\ref{S3.2}; these CMDs are modeled using maximum likelihood
techniques to extract the intensity of SF over the last 500 Myr in
\S~\ref{S3.3}.

\begin{figure}
\epsscale{1.0}
\plotone{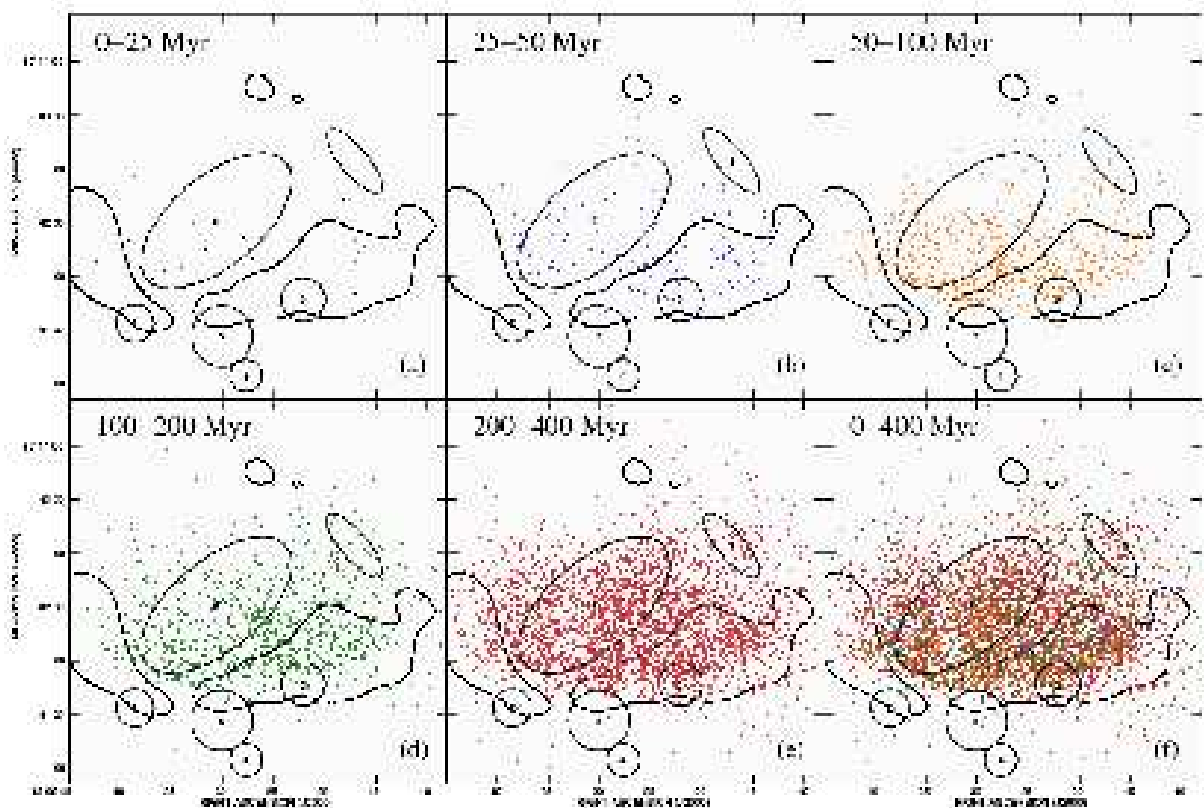}
\epsscale{1.0}
\caption{Spatial distribution of BHeB stars of different age bins, as
  labeled in the upper left corner of each panel.  Stars in each age
  bin are plotted with different colors to facilitate interpretation.
  The black contour in each panel shows the 10$^{21}$ cm$^{-2}$
  \HI\ column density contour at 20\arcsec\ resolution.  The numbered
  regions are the six \HI\ holes identified in Paper I that fall
  completely within the {\it HST} field of view.  Note the large
  increase in the number of BHeB stars from 0 to 200 Myr; BHeB stars
  permeate the entire disk at look back times of 100-400 Myr.}
\label{figcap3}
\end{figure}

We begin to explore the rich information in the global CMDs by
extracting the ages of BHeB stars in 5 coarse age bins: 0-25 Myr
(though note that the BHeB sequence is insensitive to ages $\lsim$10
Myr), 25-50 Myr, 50-100 Myr, 100-200 Myr, and 200-400 Myr.  Since the
positions of these stars are known, we can then plot the location of
BHeB stars in each age bin to allow a simple snapshot of the patterns
of SF within DDO\,165 over the last few hundred Myr.
Figure~\ref{figcap3} shows this plot; the positions of BHeB stars of
various ages are color-coded for clarity.  Overlaid in black on each
panel is the \HI\ column density contour at the 10$^{21}$ cm$^{-2}$
level, as well as the locations of the six \HI\ holes identified in
Paper I that fall completely within the {\it HST}/ACS field of view
(see more detailed discussion below).

Two important general features of the SF patterns in DDO\,165 are
apparent from Figure~\ref{figcap3}.  First, the total number of BHeB
stars rapidly increases when considering successively older age bins.
276 BHeB stars are measured in the 25-50 Myr time frame; this
increases to $>$1,100 stars with ages between 50 and 100 Myr.  This is
partially due to IMF effects, but is also suggestive of a substantial
SF event during the past 100 Myr, with more stars forming per unit
area around 100 Myr ago than at the present time.  Second, in all
epochs older than $\sim$100 Myr, SF pervades the entire disk of
DDO\,165.  This supports the spatially and temporally extended burst
scenario described above.  Figure~\ref{figcap3} demonstrates that
DDO\,165 has a complex history of SF even over the most recent 500
Myr.

From Figure~\ref{figcap2}, the tip of the red giant branch (TRGB)
occurs at m$_{\rm I}$ $=$ 24.25\,\pom\,0.1.  Assuming M$_{\rm I}$ =
$-$4.0$\pm$0.10 at the TRGB \citep{lee93,madore95,bellazzini01a} and
E(B$-$V) $=$ 0.024 mag \citep{schlegel98}, the distance of DDO\,165
derived from our {\it HST} data is in good agreement with the value
found by \citet{karachentsev02} and adopted in Paper I, D $=$
4.47\,\pom\,0.2 Mpc.  We thus retain their distance estimate throughout
the present work.

\subsection{Color-Magnitude Diagrams of Stars in Selected Regions}
\label{S3.2}

\begin{figure}
\epsscale{1.0}
\plotone{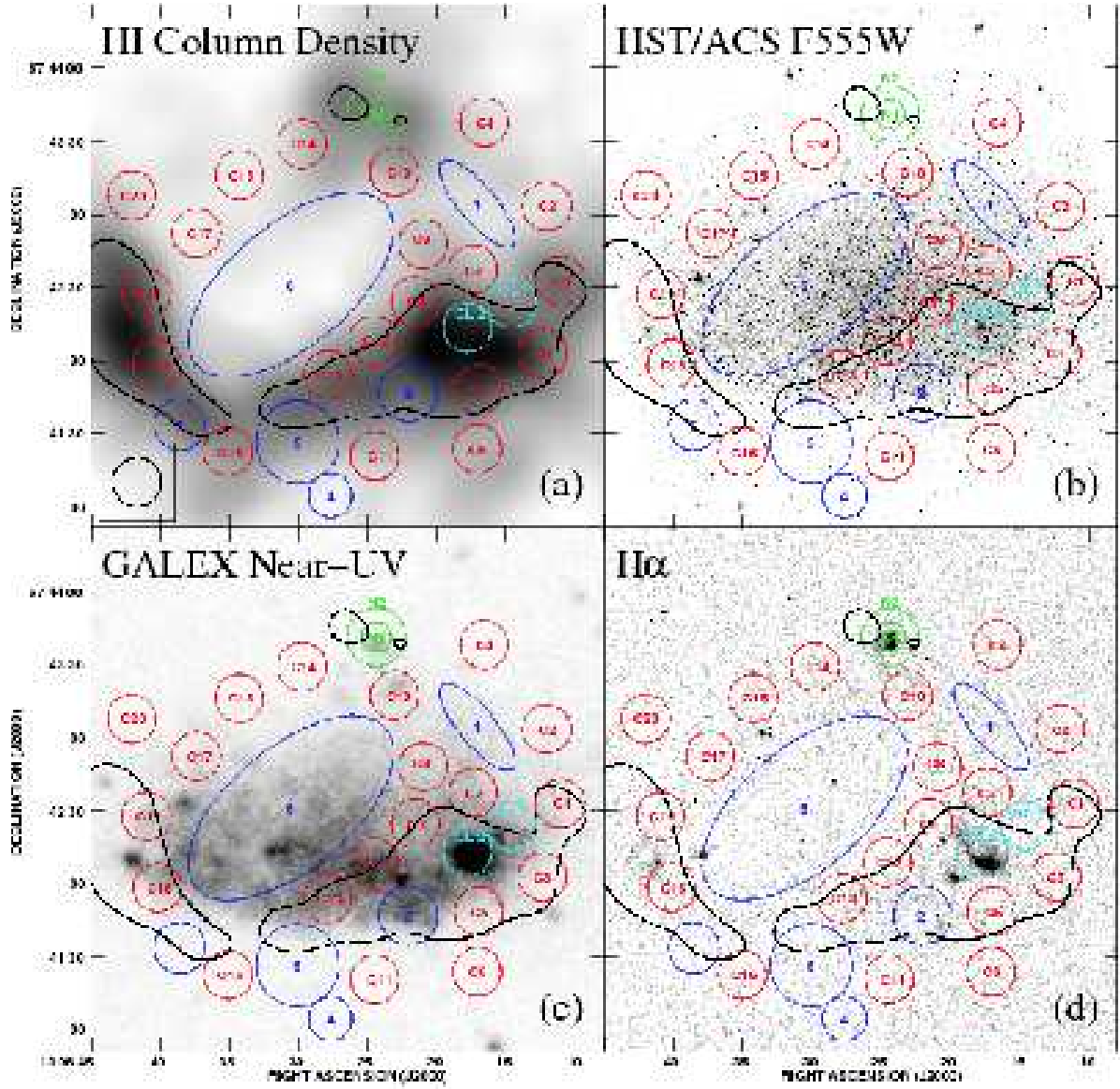}
\epsscale{1.0}
\caption{Localized regions inside which photometry is extracted,
  superimposed on the \HI\ column density image (a), the {\it
    HST}/F555W image (b), the {\it GALEX} near-UV image (c), and the
  continuum-subtracted \halpha\ image (d).  The black contour in each
  panel shows the 10$^{21}$ cm$^{-2}$ \HI\ column density contour at
  20\arcsec\ resolution.  The regions indicated are holes/shells found
  in the \HI\ (blue; see paper I), the northern region (green), visual
  clusters (cyan), and control fields (red).  Note the variety of
  stellar densities, UV and \halpha\ surface brightnesses probed by
  these regions.}
\label{figcap4}
\end{figure}

Figure~\ref{figcap4} shows four images corresponding to the {\it
  HST}/ACS field of view: \HI\ column density (see Paper~I), {\it HST}
V-band, {\it GALEX} near-UV \citep{gildepaz07}, and
continuum-subtracted \halpha\ (see Paper~I and {Kennicutt
  \etal\ 2008}\nocite{kennicutt08}).  Overlaid on each panel is the
\HI\ column density contour at the 10$^{21}$ cm$^{-2}$ level.  The
largest \HI\ column density (in the SW) is coincident with a high
surface brightness stellar cluster in the UV image, as well as with
the brightest \halpha-emitting region.  The \HI\ peak in the SE region
also has coincident UV and diffuse \halpha\ emission.  The northern
\HI\ feature is coincident with a compact \HII\ region and with a
stellar association visible in both the {\it HST} and the {\it GALEX}
images.

Paper I identified eight holes in the \HI\ distribution of DDO\,165
using radius-velocity and/or position-velocity analyses.
Figure~\ref{figcap4} shows six of those regions in blue, using the
same numbering scheme as Paper I (Holes 3 and 8 from Paper I fall
completely or partially outside the {\it HST}/ACS field of view);
Table~\ref{t1} summarizes the properties of these structures from
Paper I.  We seek to probe the SFHs of the stellar populations within
these regions and to compare them with the SFHs extracted in other
areas of DDO\,165.

For comparison, we extracted the SFHs of 23 other regions
within DDO\,165.  As shown in Figure~\ref{figcap4}, these regions
sample a wide range of stellar and \HI\ surface densities,
\halpha\ and UV surface brightnesses, and localized \HI\ dynamics
(compare with Paper~I).  We explicitly study the stellar populations
associated with the northern \HI\ component (shown in green in
Figure~\ref{figcap4} as ``N1'' and ``N2'', and referred to as
``Northern 1'' and ``Northern 2'' below) and with two regions
containing slightly resolved stellar clusters (shown in cyan in
Figure~\ref{figcap4} as ``CL 1'' and ``CL 2'', and referred to as
``Cluster 1'' and ``Cluster 2'' below).  Exterior to these regions and
the \HI\ holes, we then selected 20 control fields (shown in red in
Figure~\ref{figcap4} as ``C1''--``C20'', and referred to as ``Control
1'' through ``Control 20'' below).  All regions except for Northern 1
and Northern 2 have a diameter of 20\arcsec\ (the size of the
\HI\ beam used in this analysis; 20\arcsec\ $=$ 430 pc at our adopted
distance); this choice is necessarily a compromise between region size
and number of stars in the resulting CMDs.  The 20\arcsec\ size
provides of order 1,000 stars in each CMD (sometimes many more) while
allowing us to study a sizeable number of regions within DDO\,165.
The sizes of regions Northern 1 (12\arcsec\ diameter) and Northern 2
(24\arcsec\ diameter) are guided by eye to encompass the stellar
association in the {\it HST} image.

\begin{figure}
\epsscale{0.9}
\plotone{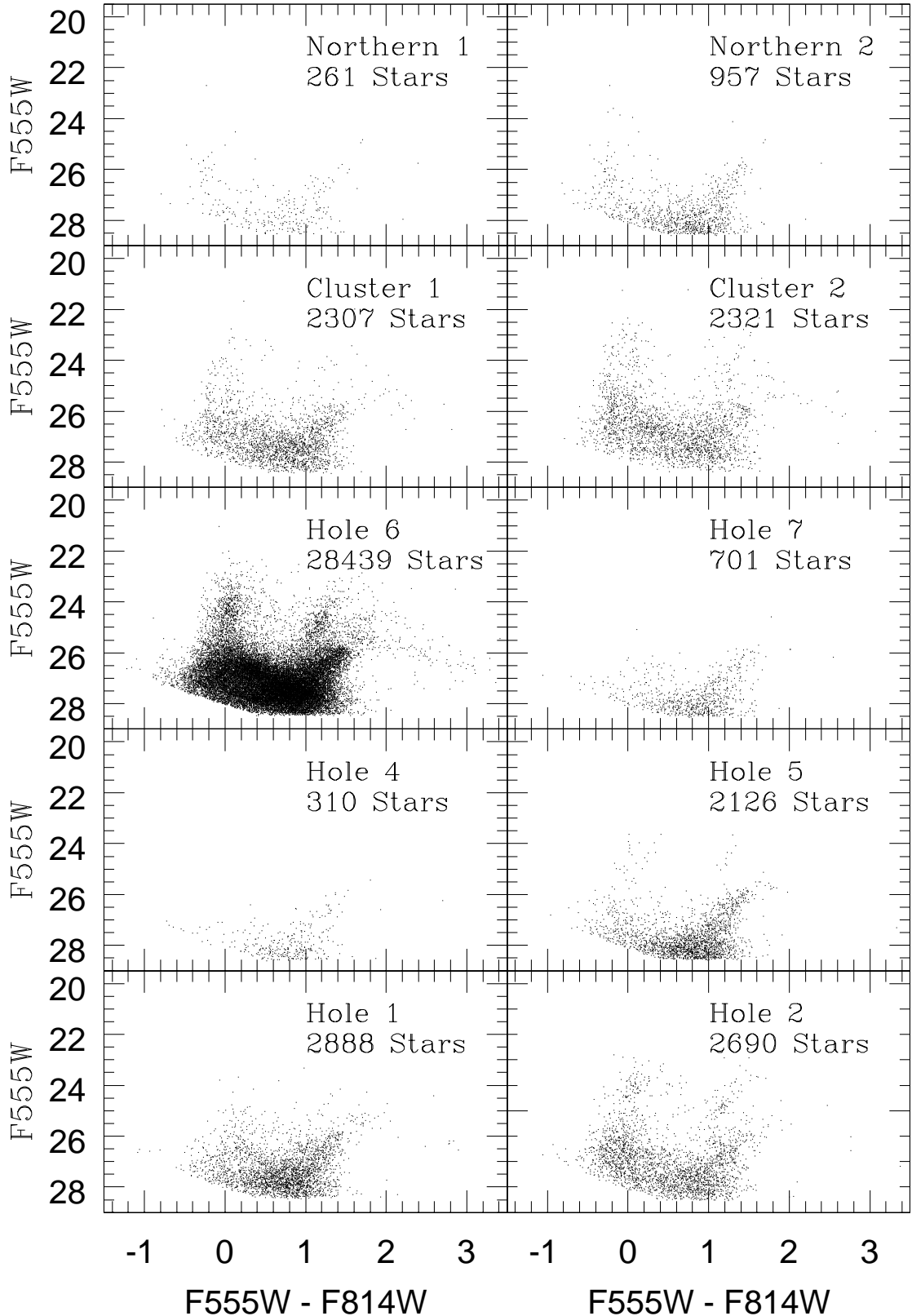}
\epsscale{1.0}
\caption{CMDs of \HI\ holes (see Paper I), regions containing visually
  identified clusters, and the kinematically distinct northern region
  (extracted over 12\arcsec\ and 24\arcsec\ circular areas in Northern
  1 and Northern 2, respectively).}
\label{figcap5}
\end{figure}

\begin{figure}
\epsscale{0.9}
\plotone{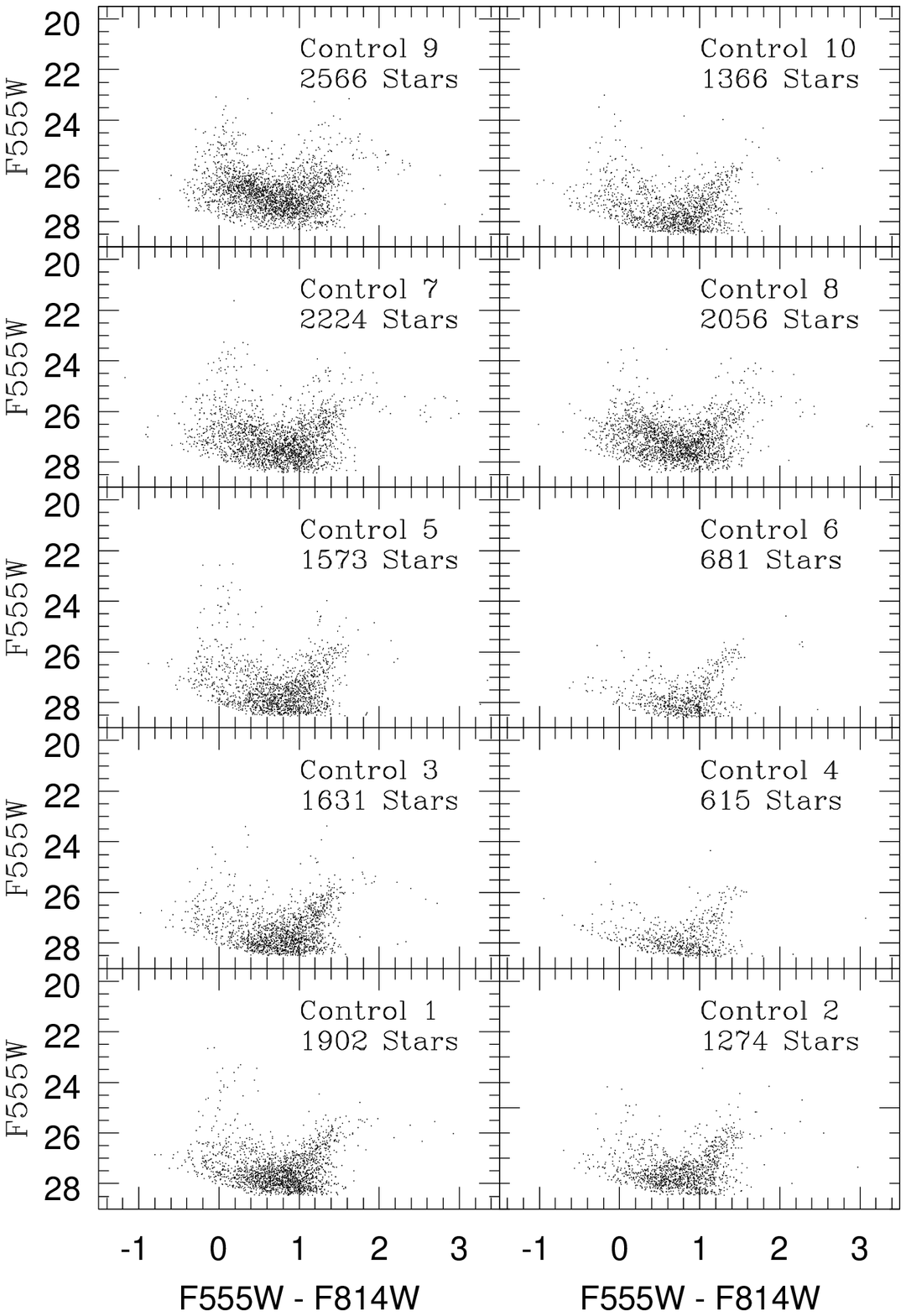}
\epsscale{1.0}
\caption{Same as Figure~\ref{figcap5}, for 
control regions 1 through 10.}
\label{figcap6}
\end{figure}

\begin{figure}
\epsscale{0.9}
\plotone{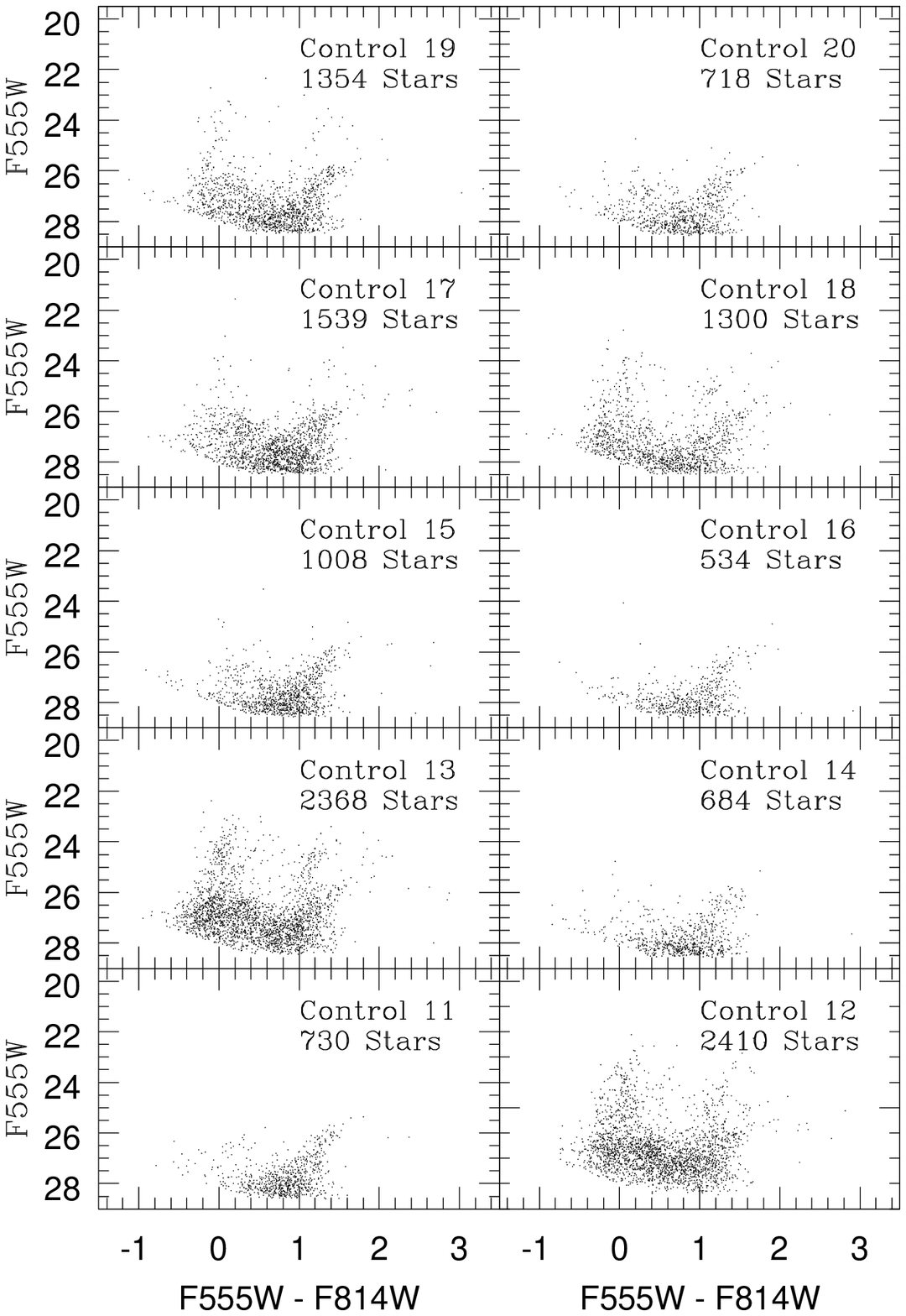}
\epsscale{1.0}
\caption{Same as Figure~\ref{figcap5}, for 
control regions 11 through 20.}
\label{figcap7}
\end{figure}

Table~\ref{t2} summarizes the names, positions and sizes of each of
these regions.  We extracted the photometry for those stars that lie
within each region; the number of stars in each region, shown in
Table~\ref{t2}, ranges from 261 (Northern 1) to 28,439 (Hole 6).  These
stars are used to construct the CMDs of each individual region shown
in Figures~\ref{figcap5} (\HI\ holes, cluster regions, northern
region), \ref{figcap6} (Control 1--10), and \ref{figcap7}
(Control 11--20).  The RGB is easily identified in all
panels; the blue plume region, however, shows pronounced variations
from one region to the next.  These variations imply significant
variations as a function of time of the locations of active SF
throughout DDO\,165.

\subsection{Recent Star Formation Histories}
\label{S3.3}

The global and local CMDs described in \S~\ref{S3.1} and \S~\ref{S3.2}
are modeled to obtain a best-fit SFH using a sophisticated maximum
likelihood approach.  We refer the reader to \citet{dolphin02} for
details, and provide a brief synopsis of the technique here.  The
stellar evolution models of \citet{marigo08} are used to construct
synthetic CMDs, which are compared with the observed CMDs using a
maximum likelihood approach.  Given the large number of factors that
contribute to the CMD of a composite stellar population (e.g., initial
mass function, chemical evolution and SFR as functions of time, binary
fraction, distance, crowding, foreground and differential extinction,
and photometric completeness limits), certain properties are held
fixed in the fitting routines.  As in \citet{weisz08}, for this
analysis of DDO\,165, the binary fraction (0.35), the IMF properties
(standard power-law, x = $-$2.30, from 0.1 to 100 \msun), the distance
(4.47 Mpc), the foreground extinction (A$_{\rm F555W}$ = 0.080;
{Schlegel \etal\ 1998}\nocite{schlegel98}), and the photometric
completeness limits (m$_{\rm F555W}$ = 28.0, m$_{\rm F814W}$ = 27.2)
are held constant.  The program is then allowed to search other
parameters to obtain the best fit, including internal extinction,
metallicity per time bin, and the SF rate (SFR) per time bin.  The
metallicity is constrained to increase monotonically with time.

The results of this fitting routine provide the rate of SF and the
chemical evolution as functions of time. As mentioned above, here we
concentrate on the most recent 500 Myr, during which we have maximum
temporal resolution via the well-populated BHeB sequence, and during
which the chemical evolution of DDO\,165 is minimal (e.g., {Weisz
\etal\ 2008}\nocite{weisz08} finds a change of 0.1 dex in
metallicity between 0 and 1 Gyr).  Errors on the SFR as a function of
time are quantified by adding the systematic uncertainties from the
stellar evolution isochrones and the statistical uncertainties from
Monte Carlo tests in quadrature.  Extensive discussion of the error
budgets can be found in \citet{dolphin02} and \citet{weisz08}.

\subsubsection{The Global Recent Star Formation History}
\label{S3.3.1}

\begin{figure}
\epsscale{1.0}
\plotone{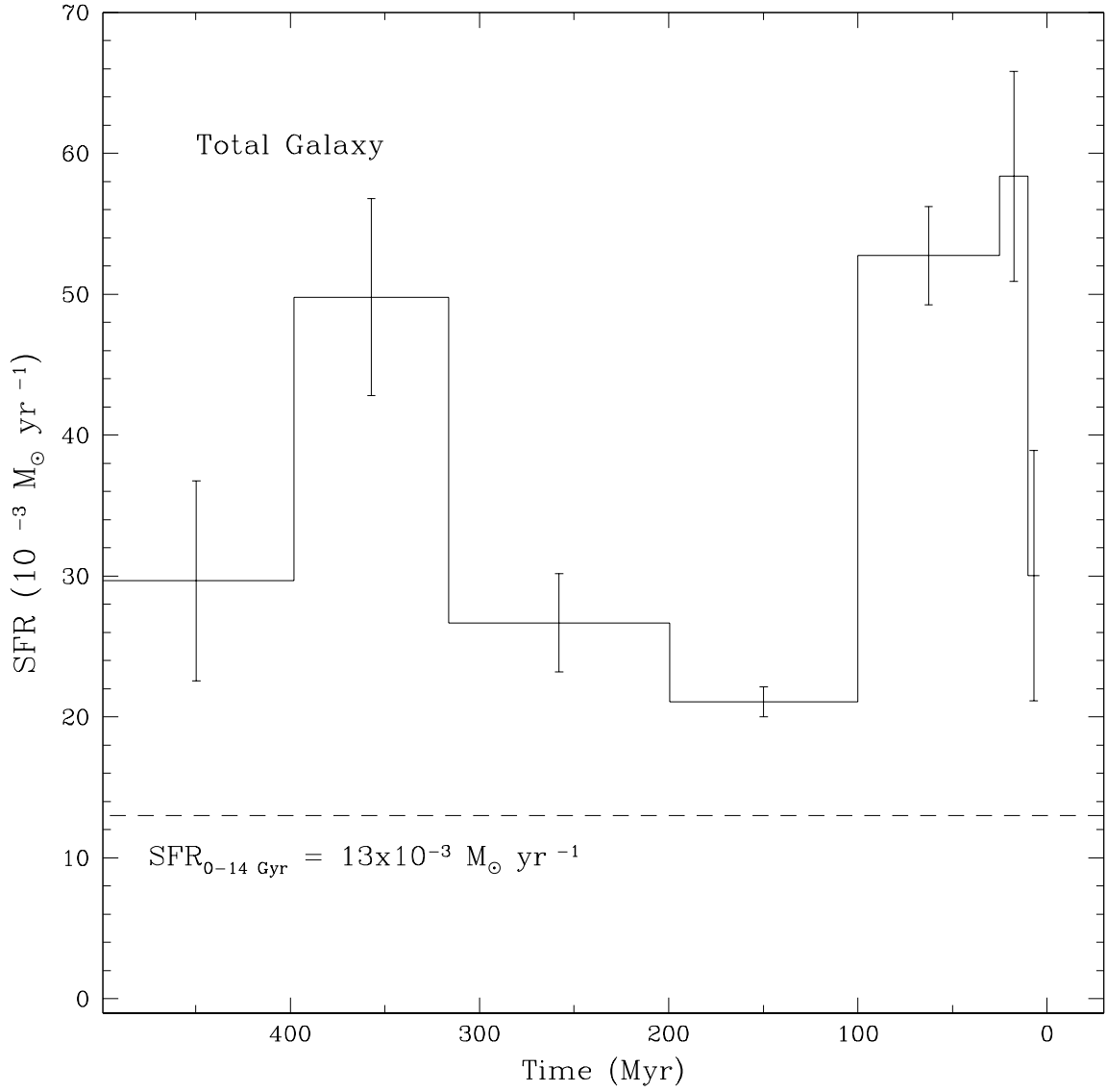}
\epsscale{1.0}
\caption{Global SFH over the most recent 500 Myr; lookback time
  increases to the left.  The dotted line shows the lifetime average
  SFR from \citet{mcquinn10}.}
\label{figcap8}
\end{figure}

Applying the CMD fitting routines described above to the global CMD
shown in Figure~\ref{figcap2} yields the SFH shown in
Figure~\ref{figcap8}.  Note that the time binning is nonlinear, with
finer temporal resolution available for young stars in the upper BHeB
and MS.  The errors plotted include both systematic and statistical
uncertainties.

As noted above and discussed in detail in \citet{mcquinn10}, DDO\,165
hosts one of the longest-duration starbursts known to date ($>$1 Gyr).
Figure~\ref{figcap8} verifies this temporally extended burst by
comparing the recent SFR with SFR$_{\rm 0-14\, Gyr}$ (dashed line),
the average SFR of the entire galaxy averaged from 0 to 14 Gyr.  Note
that the SFR is higher throughout the most recent 500 Myr compared to
the lifetime average SFR, indicating an ongoing starburst event.

Certain interesting features of the global SFH are apparent.  First,
the SFR varies by a factor of $\sim$2 during the last 500 Myr: during
the intervals from $\sim$100 Myr to 25 Myr, and again between
$\sim$400 and 300 Myr, the SFR is significantly higher than the
average over the last 500 Myr.  Second, the SFR drops significantly
between 25 Myr and the present time; this is in agreement with the
lack of bright \halpha\ emission in DDO\,165 (see discussion in Paper
I and Figure~\ref{figcap4}).  The stars that populate these recent
time bins are the most luminous BHeB and MS stars; if they were
present in large numbers in DDO\,165, they would be prominent in the
global CMD.  The lower SFR over the last 25 Myr is thus interpreted as
a significant decrease in the total SFR in DDO\,165.  Whether this
current drop in SFR corresponds to the ultimate truncation of the
Gyr-length burst is not discernible from these data.

\subsubsection{Recent Star Formation Histories: Localized or Global Events?}
\label{S3.3.2}

\begin{figure}
\epsscale{1.0}
\plotone{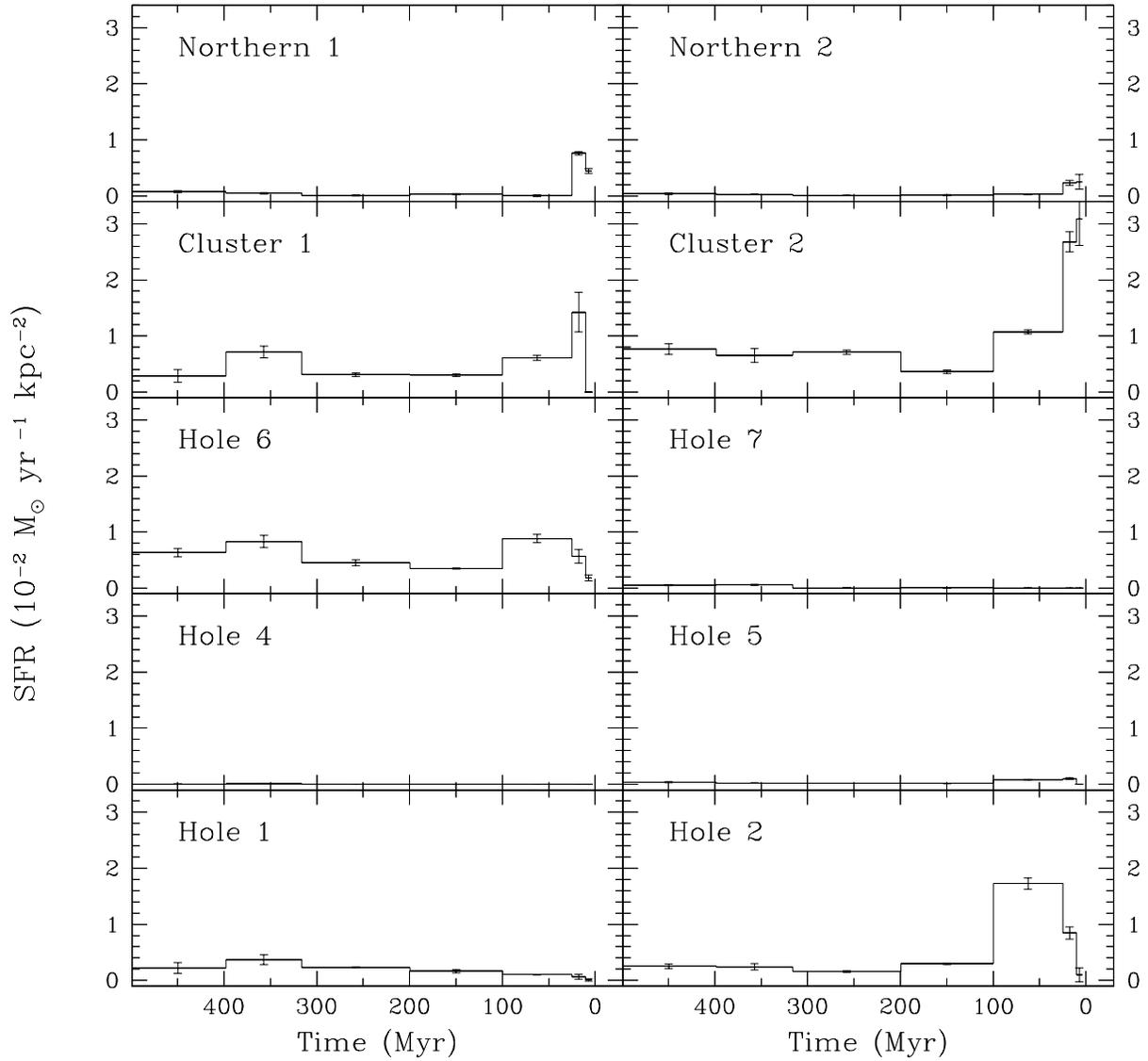}
\epsscale{1.0}
\caption{Normalized SFHs of localized regions (see Table~\ref{t2});
  the CMDs of these fields are shown in Figure~\ref{figcap5}.
  Lookback time increases to the left.}
\label{figcap9}
\end{figure}

\begin{figure}
\epsscale{1.0}
\plotone{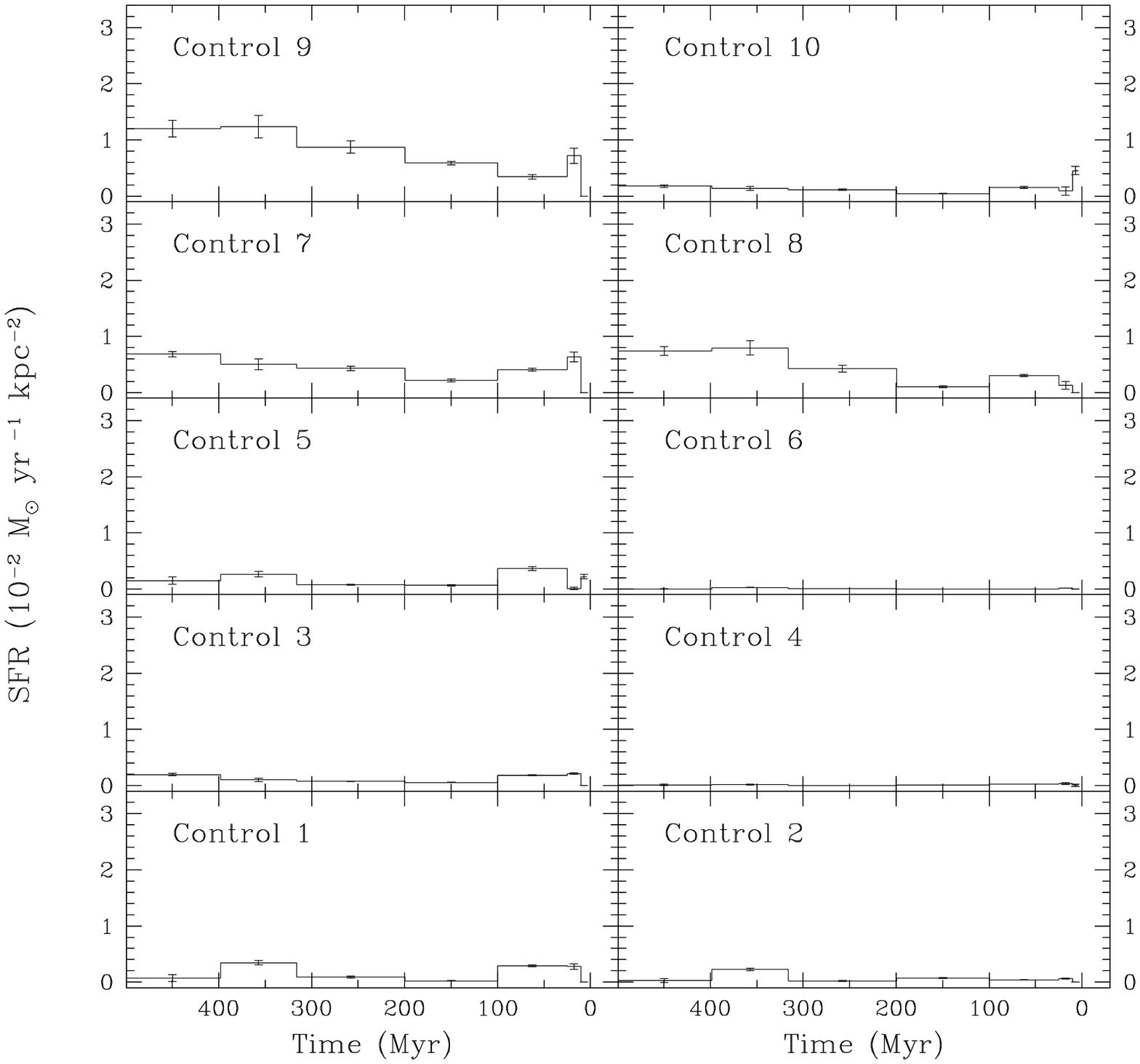}
\epsscale{1.0}
\caption{Same as Figure~\ref{figcap9}, but for Control 1--10; 
 the CMDs of these fields are shown in Figure~\ref{figcap6}.}
\label{figcap10}
\end{figure}

\begin{figure}
\epsscale{1.0}
\plotone{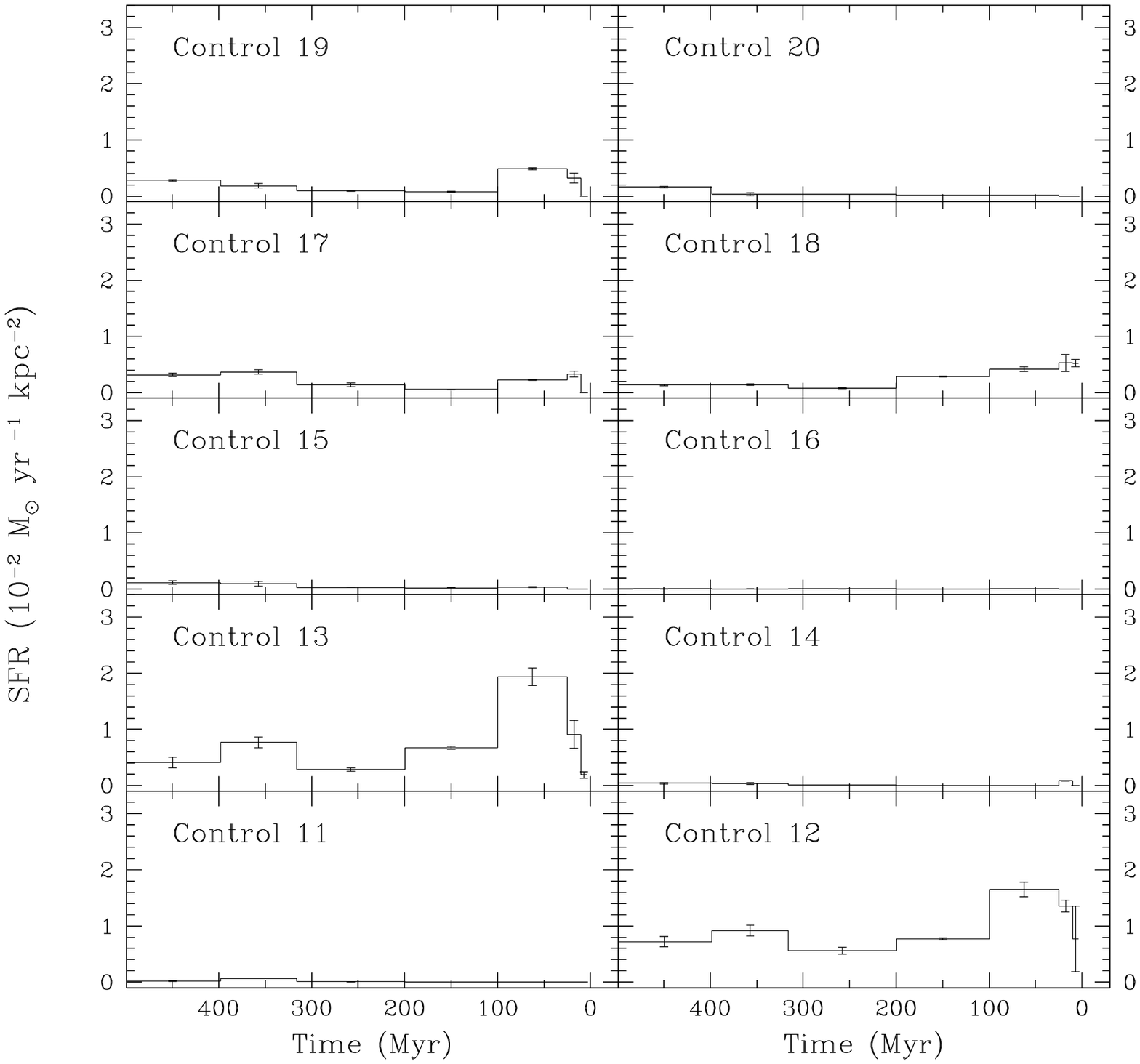}
\epsscale{1.0}
\caption{Same as Figure~\ref{figcap9}, but for Control 11--20; 
 the CMDs of these fields are shown in Figure~\ref{figcap7}.}
\label{figcap11}
\end{figure}

Figures~\ref{figcap9}, \ref{figcap10}, and \ref{figcap11} present the
SFHs of the selected regions in DDO\,165 shown in Figure~\ref{figcap4}
and discussed in \S~\ref{S3.2}; each region is normalized to show SFR
per unit time per unit area.  The normalized recent SFHs span an
intensity range of $\sim$3 over the last 500 Myr in these regions.
There are clear variations in normalized SFRs from one field to the
next: some regions have constant SFRs over the last 500 Myr (e.g.,
Hole 1, Control 5), while others have dramatic variations in their
SFRs during certain intervals (e.g., Northern 1, Hole 2, Cluster 2,
Control 12, Control 13).  Comparing with the multiwavelength
properties shown in Figure~\ref{figcap4}, the localized SFHs within
DDO\,165 arise from a wide variety of local conditions.

Two important time intervals are apparent from the localized SFHs:
400-300 Myr and 100-25 Myr.  These epochs are identified as periods of
enhanced local normalized SFR compared to the 500 Myr average.  In the
more ancient event, the global SFR increases by $\sim$50\% compared to
the 500 Myr average.  A corresponding increase in normalized SFR is
seen in \HI\ Holes 1 and 6, Cluster 1, and Control 1, 2, 5, 8, 9, 11,
12, 13, and 17.  These regions are distributed over a large fraction
of the stellar body of DDO\,165, indicating that a global SF event
took place during this period.  The second period of enhanced SF is
100-25 Myr; the global SFR again increases by $\sim$50\%.  A
corresponding elevated SFR is seen in regions located throughout
DDO\,165: Holes 2, 5, and 6, Clusters 1 and 2, the northern
\HI\ component (Northern 1 and Northern 2), and Control 1, 3, 5, 7, 8,
10, 12, 13, 17, 18 and 19.  The fact that this more recent SF event
occurred in widespread regions of the galaxy (compare, for example,
the locations and SFHs of Northern 1, Control 1 and 18) suggests a
second global SF event.  The rapid truncation of the global SFR, as
well as the decrease in most normalized SFRs, suggests that this event
ended as quickly as it began.

We note with interest that the 400-300 Myr interval identified in this
SFH analysis agrees with two other relevant timescales for DDO\,165.
The first is the well-known three-body tidal interaction between
M\,81, M\,82, and NGC\,3077 that occurred $\sim$300 Myr ago
\citep{vanderhulst79,yun94}.  However, since no direct evidence is
seen in our \HI\ data for a recent tidal interaction (see Paper I),
nor does any kinematic model of the M\,81 group suggest that DDO\,165
was directly involved in the M\,81 - M\,82 - NGC\,3077 interaction
\citep[e.g.,][]{thomasson93}, this timescale may be coincidence.  The
second and perhaps more important timescale is t$_{\rm evo}$, the time
required to remove evidence of correlated stellar structure.
\citet{bastian11} has analyzed the evolution of stellar structures in
various M\,81 group dwarf galaxies, and finds that stellar structures
survive for at least 350 Myr in DDO\,165.  The long t$_{\rm evo}$ for
DDO\,165 (compare to $>$80 Myr for the SMC and $>$100 Myr for
NGC\,2366; see {Gieles \etal\ 2008}\nocite{gieles08} and {Bastian
  \etal\ 2011}\nocite{bastian11}) guarantees that the recent SFHs
within the various regions of the galaxy can be reliably compared with
the current \HI\ properties of the same regions; we discuss this point
further in \S~\ref{S4.3} below.

\subsubsection{Comparison to Individual Star Clusters}
\label{S3.3.3}

We have identified 20 individual stellar clusters to examine any temporal
consistencies with the SFHs of the fields presented so far.  The
clusters were identified by inspection of a F555W/F814W color {\it
  HST} image.  The masses and ages of the clusters tabulated in
Table~\ref{t3} were determined from fitting broadband BVRI
ground-based photometry to single stellar population models
\citep{bertelli94,marigo08}.  The BVRI data are part of the Spitzer
Infrared Nearby Galaxies Survey (SINGS; {Kennicutt
  \etal\ 2003}\nocite{kennicutt03}); images were acquired with the
{\it KPNO 2.1-m} telescope.

\begin{figure}
\epsscale{1.0}
\plotone{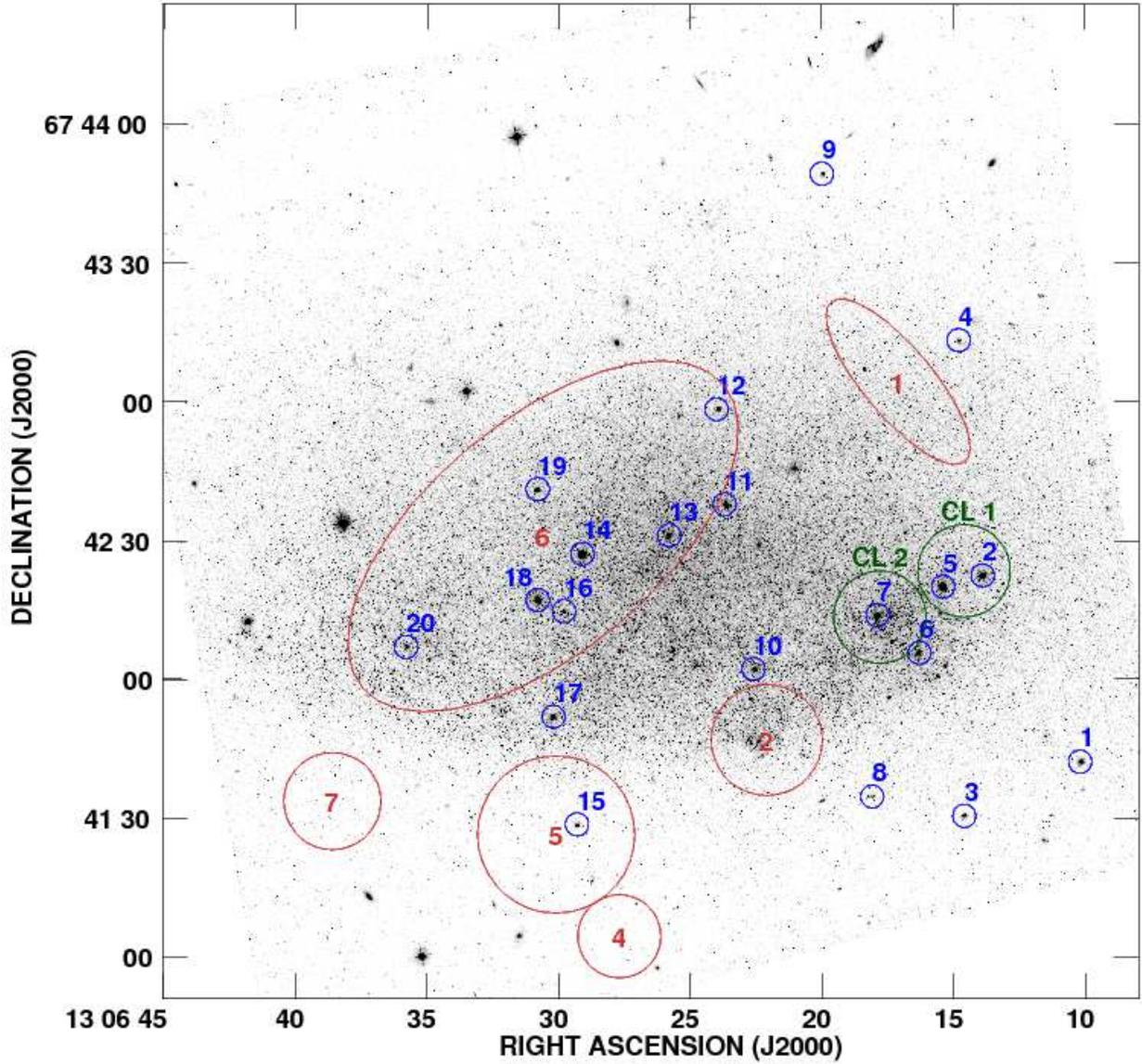}
\epsscale{1.0}
\caption{{\it HST}/F555W image of DDO\,165, overlaid with the
  positions of compact stellar clusters (blue; see also Table~\ref{t3}
  and discussion in \S~\ref{S3.3.3}), the six \HI\ holes (red; see
  also Table~\ref{t1}), and the Cluster 1 and Cluster 2 regions
  (green; see Table~\ref{t2} and compare to Figure~\ref{figcap4}).}
\label{figcap12}
\end{figure}

Figure~\ref{figcap12} shows the spatial relationship of these
individual clusters to the \HI\ holes and Cluster regions identified
above.  Note that we use lower case ``c'' labels for individual
clusters in Table~\ref{t3}.  Holes 5 and 6 both contain stellar
clusters, as do the regions Cluster 1 and Cluster 2.  Conversely,
there are 4 holes that do not contain readily identifiable
clusters. Also of interest is that the strongest emission in
\halpha\ and UV comes from the region inhabited by the individual
clusters c2, c5, c6, and c7.  

\begin{figure}
\epsscale{1.0}
\plotone{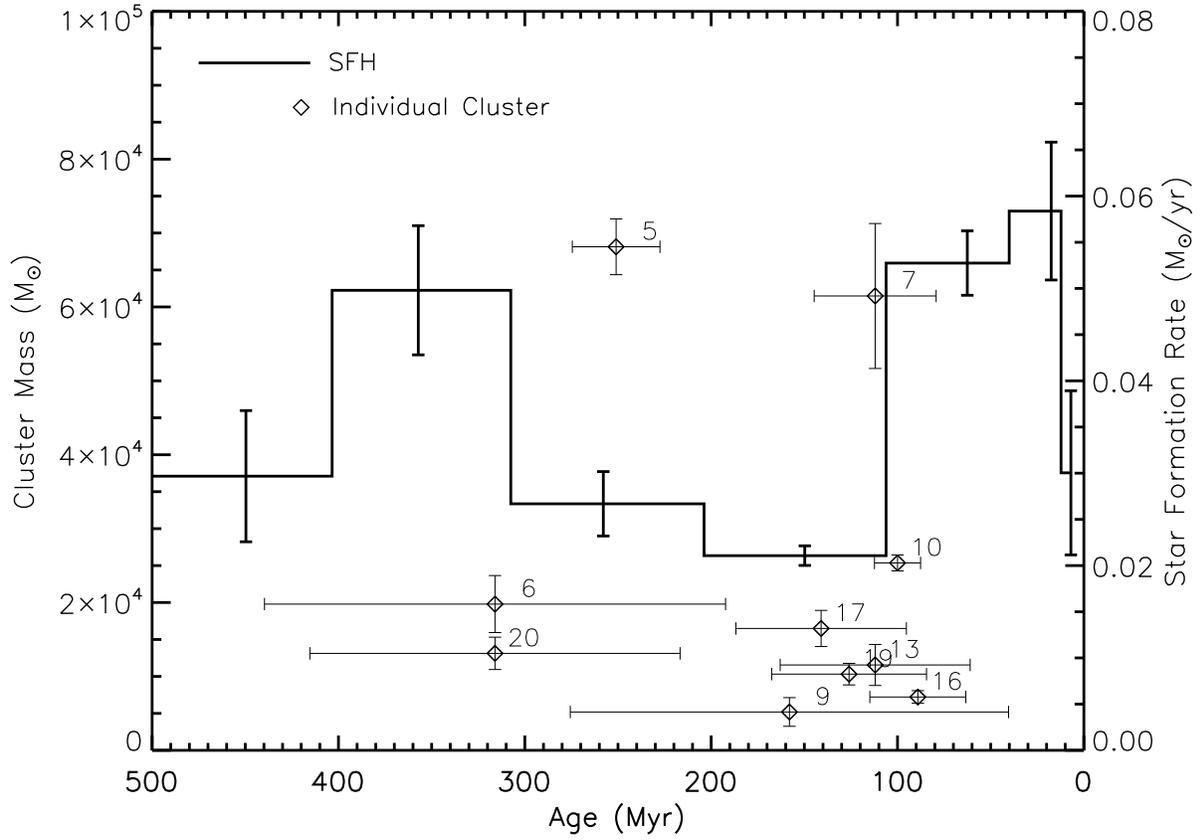}
\epsscale{1.0}
\caption{Comparison of the global SFR (solid line; see
  discussion in \S~\ref{S3.3.1}) with the masses and ages of individual
  stellar clusters (see discussion in \S~\ref{S3.3.3}). Note the
  concentration of clusters with ages near 100 Myr, when the total
  SFR increases.}
\label{figcap13}
\end{figure}

Next we examine the cluster age properties in relation to the global
SFH of DDO\,165 over the last 500 Myr.  Figure~\ref{figcap13} shows
the global SFH (identical to the SFH shown in 
Figure~\ref{figcap8}), overplotted with the masses and ages of
individual stellar clusters.  The most notable feature is the temporal
concentration of clusters (c7, c9, c10, c13, c16, c17, and c19) near
100 Myr.  This age corresponds to the increase in global SFR at the
onset of the 100-25 Myr SF event discussed above.  Note also that the
7 clusters with ages of $\sim$100 Myr are scattered throughout the
entire disk of DDO\,165.  Taken together, these properties suggest
that the formation of stellar clusters is directly linked to the increase 
in global SFR in DDO\,165.

Based on the results for DDO\,165 presented here, the connection
between \HI\ holes and stellar clusters appears to be somewhat
ambiguous: \HI\ structures are identified both with and without
interior stellar clusters.  However, it is important to note that all
of the observed clusters have ages $\gsim$90 Myr (see Table~\ref{t3}).
This suggests that we are observing only those associations that have
survived the well-documented ``infant mortality'' phase (e.g., {Lada
  \& Lada 2003}\nocite{lada03}; {Fall \etal\ 2005}\nocite{fall05};
{Chandar \etal\ 2006}\nocite{chandar06}; and references therein).  If
the clustered SF in DDO\,165 has proceeded in a manner similar to that
observed in other systems, it seems probable that many young clusters
have been destroyed within the past $\sim$100 Myr.

It is interesting to note that the relation between total SFR
(clusters plus field stars) and cluster formation rate during periods
of heightened SF activity appears to be linear in log-log space
\citep{goddard10,silva10}.  To date, this relation has been described
in the literature for 10 galaxies, none of which are dwarfs.  The
results above for DDO\,165 appear to agree with this log-log linear
relationship between total SFR and cluster formation rate.  Further
work to solidify this realtion in low-mass galaxies would be very
instructive.

\section{Probing the Stellar-ISM Interaction in DDO\,165}
\label{S4}

\subsection{Patterns of Recent Star Formation Versus \HI}
\label{S4.1}

The preceding discussion highlights that SF has pervaded the disk of
DDO\,165 over the past 500 Myr.  Figure~\ref{figcap3} shows that over
this time interval, the BHeB stars are distributed more or less
uniformly throughout the inner \HI\ disk.  However, note that the
truncated stellar surface brightness of the southern regions (compare
with the figures presented in Paper I) remains coherent in the BHeB
populations at all age intervals.  This result, together with the
characteristic t$_{\rm evo}$ $\simeq$ 350 Myr derived in
\citet{bastian11}, motivates our comparison of the patterns of SF over
the last 500 Myr with the current \HI\ morphology and dynamics.
Although the \HI\ data lack temporal information, the arguments for
the dynamical simplicity of dwarfs (e.g., lack of internal shear, long
dynamical timescales) presented in Paper I, and the predicted survival
timescales of \HI\ holes from simulations ($\sim$100-600 Myr; {Recchi
  \& Hensler 2006}\nocite{recchi06}), suggest that the comparison of
recent SF and present-day \HI\ properties are physically motivated in
low-mass systems.

We take advantage of the fine temporal and physical resolution in our
{\it HST} data by presenting movies of the normalized SFR.  These
movies were created using a procedure similar to that described in
\citet{dohmpalmer97} and applied by \citet{weisz09b}; we refer the
reader to the latter work for detailed discussion.  The first step is
to create a series of stellar density maps on a specified grid size.
Knowing the physical coordinates and ages of the BHeB stars, they are
binned into 5 Myr intervals and gridded in Right Ascension and
Declination.  These images are then smoothed spatially by a
17\arcsec\ Gaussian kernel (most directly comparable to the
20\arcsec\ \HI\ beam size from Paper I) and then normalized by
dividing by the total number of counts.

The second step is to create a SFH on the same temporal grid.  The
global SFH is binned with different time resolutions to account for
the degraded resolution with SFH look back time: 10 Myr for the 0-100
Myr interval, 20 Myr for the 0-200 Myr interval, and 50 Myr for the
0-500 Myr interval.  These are linearly interpolated onto the same
temporal grid as the stellar density maps (5 Myr).  The multiplicative
product of the stellar density maps and the SFH is divided by the
physical area of the field of view, thus resulting in a series of
images representing SFR per unit area, each separated by 5 Myr.

\begin{figure}
\epsscale{0.8}
\plotone{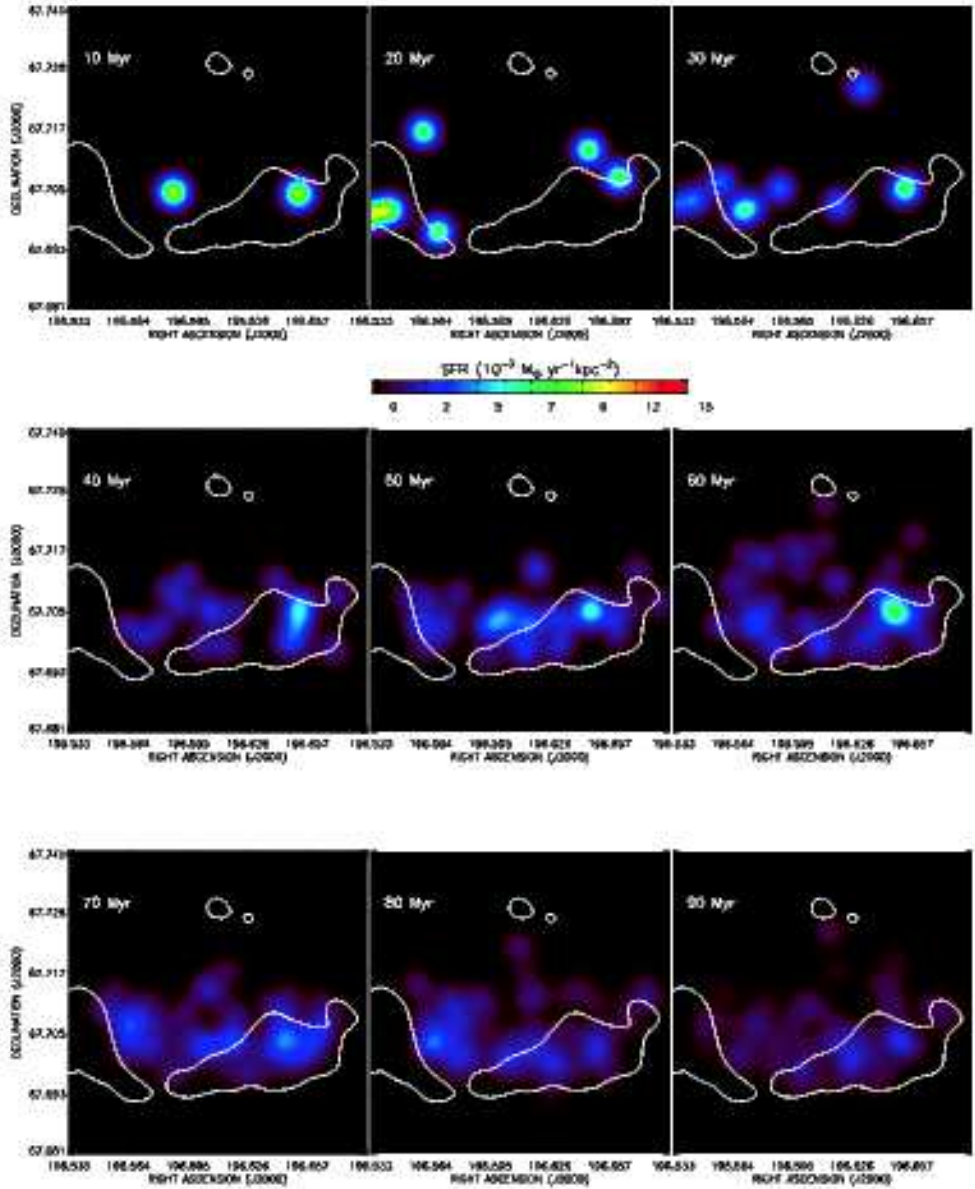}
\epsscale{1.0}
\caption{Still frames of the spatially resolved SFH in DDO\,165 over
  the last 100 Myr at 10 Myr temporal resolution.  The intensity scale
  corresponds to SFR per unit area at the adopted distance; note that
  the dynamic range is higher than in Figures~\ref{figcap15} and
  \ref{figcap16}.  The 10$^{21}$ cm$^{-2}$ \HI\ column density contour
  (20\arcsec\ resolution) is overlaid in white.  SF has been mostly
  concentrated in the southern \HI\ component during the last 100 Myr;
  the bright regions in the recent time bins correspond to one or
  a few individual massive stars.  An animation of this figure is
  available in the online version of the journal.}
 \label{figcap14}
\end{figure}

\begin{figure}
\epsscale{0.8}
\plotone{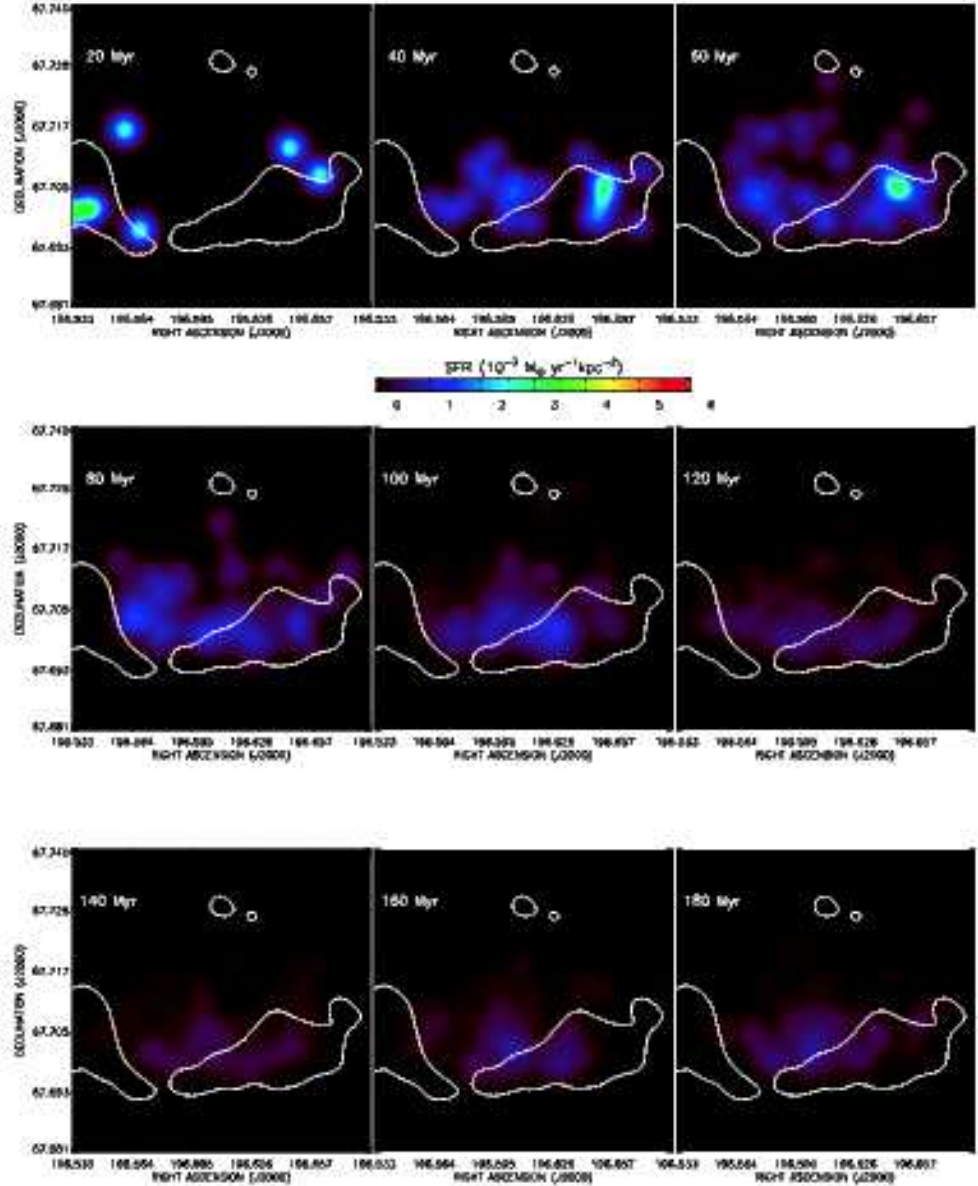}
\epsscale{1.0}
\caption{Still frames of the spatially resolved SFH in DDO\,165 over
  the last 200 Myr at 20 Myr temporal resolution.  The intensity scale
  corresponds to SFR per unit area at the adopted distance; note that
  the dynamic range is lower than in Figure~\ref{figcap14} but higher
  than in Figure~\ref{figcap16}.  The 10$^{21}$ cm$^{-2}$ \HI\ column
  density contour (20\arcsec\ resolution) is overlaid in white.  While
  less intense per unit area than in the recent 100 Myr (see
  Figure~\ref{figcap14}), the SF is more widespread throughout the
  disk over the past 200 Myr.  An animation of this figure is
  available in the online version of the journal.}
\label{figcap15}
\end{figure}

\begin{figure}
\epsscale{0.8}
\plotone{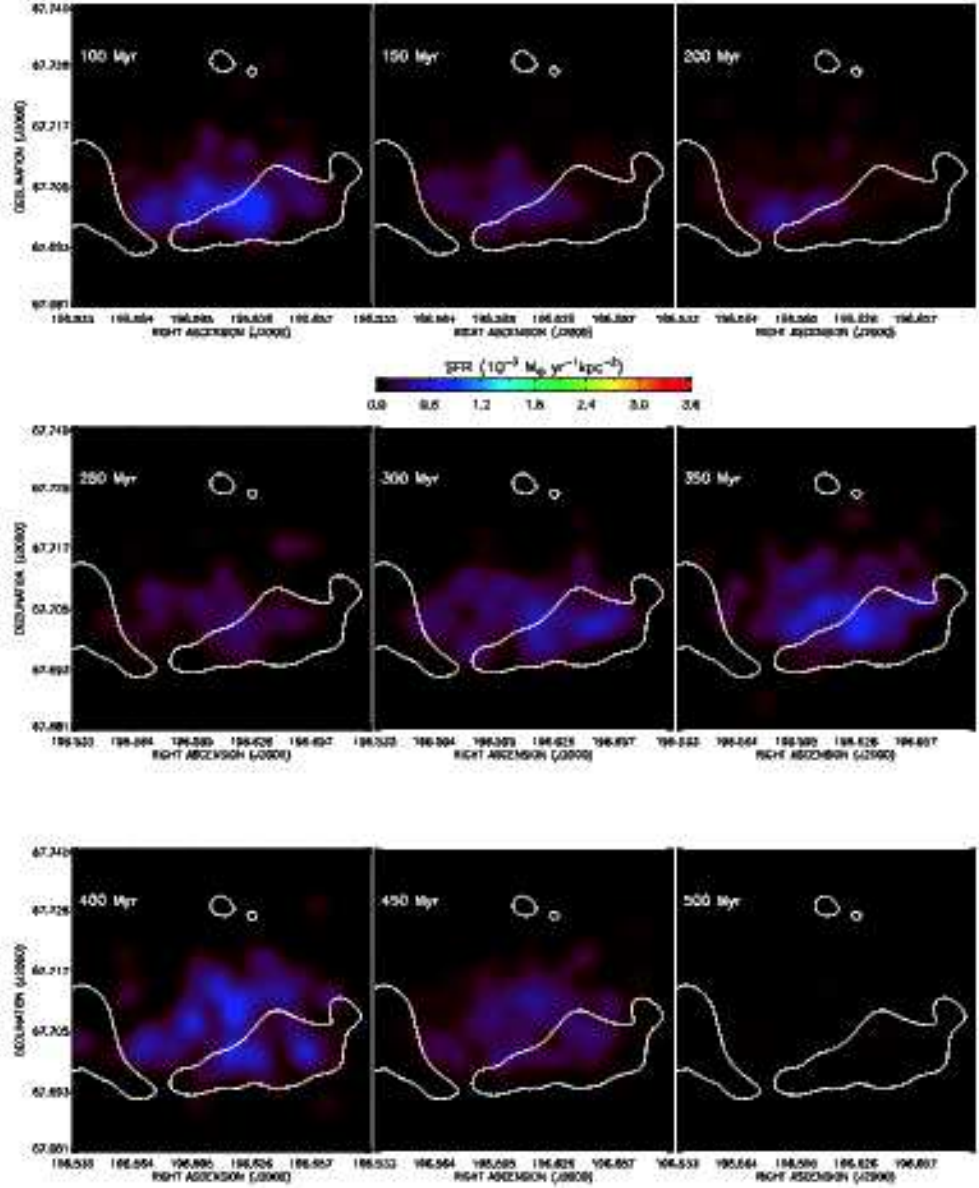}
\epsscale{1.0}
\caption{Still frames of the spatially resolved SFH in DDO\,165 over
  the last 500 Myr at 50 Myr temporal resolution.  The intensity scale
  corresponds to SFR per unit area at the adopted distance; note that
  the dynamic range is lower than in Figures~\ref{figcap14} and
  \ref{figcap15}.  The 10$^{21}$ cm$^{-2}$ \HI\ column density contour
  (20\arcsec\ resolution) is overlaid in white.  Note that most of the
  SF in the 400-500 Myr time range is concentrated in the central
  disk, close to the location of the largest \HI\ hole (Hole 6).  An
  animation of this figure is available in the online version of the
  journal.}
\label{figcap16}
\end{figure}

The resulting movies (available in the electronic version of the
journal) show the intensity of SF as a function of time and of
position, to lookback times of 100 Myr, 200 Myr, and 500 Myr,
respectively. Note that while each movie has the same time step (5
Myr), the movies extending to larger lookback times were created by
interpolation over larger time intervals and thus have a higher
relative uncertainty from one time step to the next.  For clarity,
each movie contains the same \HI\ 10$^{21}$ cm$^{-2}$ column density
contour as shown in Figure~\ref{figcap4}.  Figures~\ref{figcap14},
\ref{figcap15} and \ref{figcap16} show selected still frames from the
movies.  It is important to emphasize that these movies and images
show a SFR per unit area; high SFRs can arise from low-level SF
distributed over large areas, or from high-level SF in a small region.
Using the heightened period of SF around 100 Myr (see
Figure~\ref{figcap8}) as a guide, the SFRs per unit area in the movies
near this lookback time are at modest levels but occur over a
substantial portion of the system.

The movies highlight various important features of the recent
evolution of this intriguing system.  First, the starburst in DDO\,165
is a global event; throughout the last 500 Myr the SF has
been ongoing throughout most of the stellar component.  Second, the
period of heightened global SFR rate from 400-300 Myr is apparent as
distributed SF throughout the system and centered on the current
stellar distribution; the elevated global SFR from 100-25 Myr is
manifested as somewhat more localized regions of SF (preferentially
located in the southern region) superposed on distributed lower-level
SF.  Third, there is little SF exterior to the 10$^{21}$
cm$^{-2}$ column density contour.  This is especially evident in the
southern region as the aforementioned truncation of the stellar
population.  Finally, comparing with Figure~\ref{figcap3} shows that
SF inside the giant \HI\ hole has spanned the entire 500 Myr interval
(although only a small number of stars with ages $<$25 Myr are
present).  

In addition to the \HI\ holes and shells discussed above, Paper I
identified a wealth of high-velocity gas within DDO\,165.  Most of
this gas is located inside of the southern 10$^{21}$ cm$^{-2}$
contours shown in the movies and in Figures~\ref{figcap14},
\ref{figcap15}, and \ref{figcap16}. A careful examination of the
spatial distribution of BHeB stars in these regions
(Figure~\ref{figcap3}), the normalized SFHs of these regions (e.g.,
Clusters 1 and 2, Hole 2, Control 7, 9, 12, 13, 18), and the SFR per
unit area movies shows that there has been significant SF in the
southern \HI\ component at locations coincident with the present-day
high velocity neutral gas.  In agreement with Paper I, extended SF
events appear capable of creating incoherent but high velocity motion
of the surrounding interstellar material.  We note that coherent
\HI\ structures smaller than our physical resolution limit ($\sim$160
pc using our highest resolution \HI\ cubes) may be present in
DDO\,165.

\subsection{The Energies of \HI\ Holes}
\label{S4.2}

Paper I identified 8 holes and shells in the neutral ISM of DDO\,165.
Of those 8 structures, 5 were found to have kinematic expansion
signatures.  Table~\ref{t1} summarizes the properties of the
\HI\ holes that lie within the {\it HST} field of view.  For the three
kinematically expanding structures in Table~\ref{t1}, Paper~I made the
the simplistic assumption that each structure is created by a single
blast.  A coarse estimate of the energy required to create the
expanding \HI\ structure was then calculated via

\begin{equation}
{\rm E_{hole} = 5.3\times10^{43}}\, {\rm n_0^{1.12}}\, {\rm (\frac{d}{2})^{3.12}}\, {\rm v_{exp}^{1.4}} \ {\rm erg}
\label{eq1}
\end{equation}

\noindent where E$_{\rm hole}$ is the energy needed to create the
expanding structure in erg, n$_{\rm 0}$ is the \HI\ volume density
before the creation of the hole (Paper I assumes n$_{\rm 0}$=0.1
cm$^{-3}$), d is the diameter of the shell in parsecs, and v$_{\rm
  exp}$ is the observed expansion velocity in \kms.  These E$_{\rm
  hole}$ values are tabulated in Table~\ref{t4} (columns 2, 3, and 4).

As discussed in Paper I, this calculation is overly simplistic for a
number of reasons; we now attempt a more realistic and physically
motivated derivation of the energy requirements for the observed
\HI\ structures by modeling the mass surface density and the
\HI\ volume density.  For kinematically expanding structures, the
velocity and physical size are easily measured; the volume density,
however, is difficult to measure directly and we thus rely on a mass
model approach to obtain the behavior of n$_{\rm 0}$ as a function of
position.  For a disk of \HI\ gas, the column density (N$_{\rm HI}$),
volume density (n$_{\rm 0}$) and disk scale height ({\it h}) are
related via

\begin{equation}
n_0 = \frac{N_{HI}}{\sqrt{2\pi}h}
\label{eq2}
\end{equation}

\noindent Since the velocity field of DDO\,165 is so peculiar, Paper I
was not able to achieve an unambiguous rotation curve fit, which could
have been used to determine the scale height of the galaxy.  We thus
rely on infrared photometry to produce a stellar mass model of DDO
165.  Note that such a model implicitly assumes that the bulk of the
infrared luminosity arises from the stellar populations older than the
MS or BHeB stars; this population is assumed to retain the same mass
distribution both before and after a SF event.

\begin{figure}
\epsscale{1.0}
\plotone{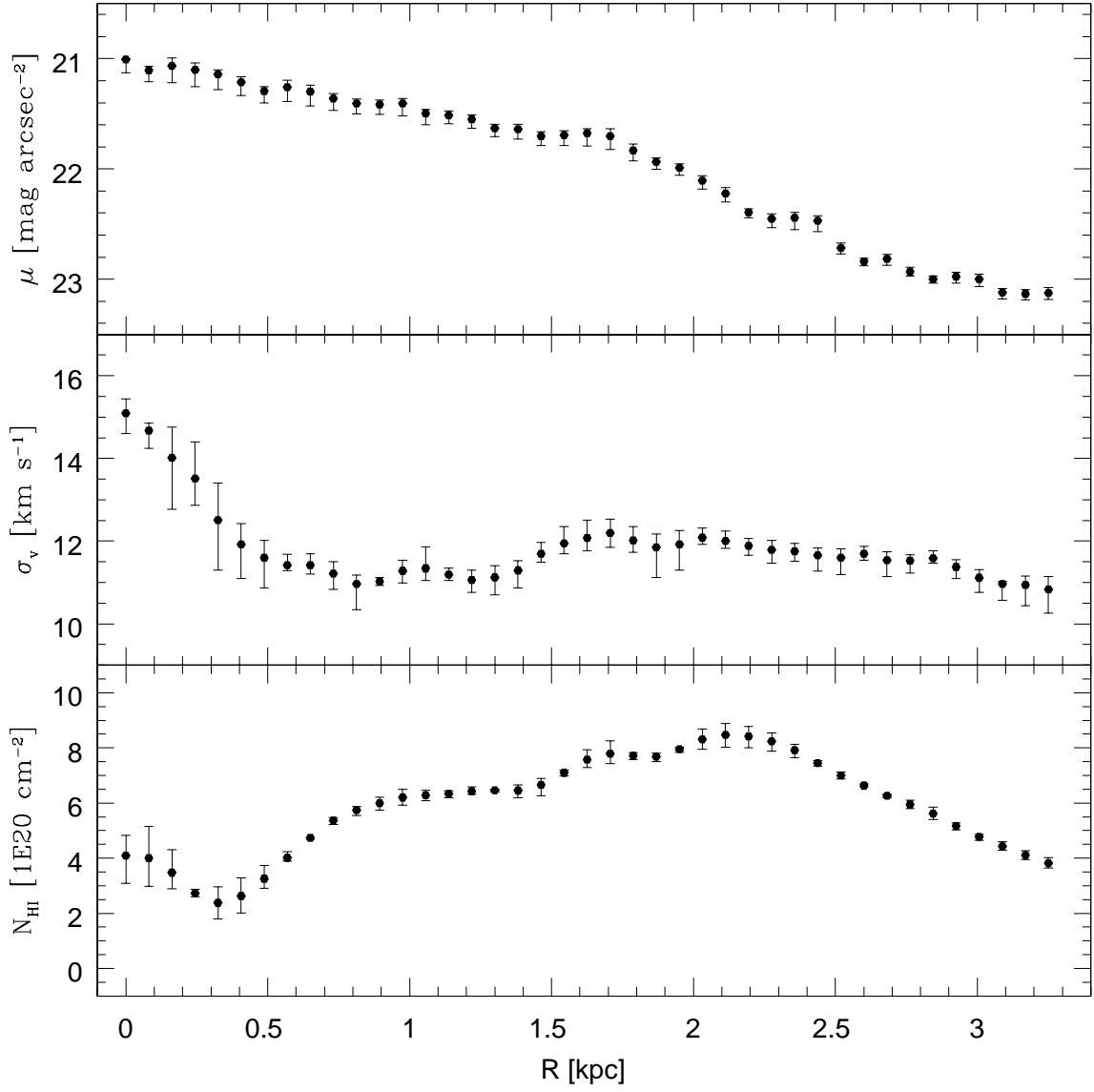}
\epsscale{1.0}
\caption{{\it Top}: Radial 3.6 $\mu$m surface brightness profile,
  derived from images acquired for the Spitzer Infrared Nearby Galaxy
  Survey \citep[SINGS;][]{kennicutt03}.  {\it Middle}: Radial velocity
  dispersion profile from the \HI\ data (see Paper I).  {\it Bottom}:
  Radial column density profile from the \HI\ data.}
\label{figcap17}
\end{figure}

We used the IRAC 3.6 $\mu$m images acquired with the {\it Spitzer
  Space Telescope} as part of SINGS (Kennicutt
\etal\ 2003\nocite{kennicutt03}; see also {Walter
  \etal\ 2007}\nocite{walter07}) to derive our stellar mass model (see
discussion in {Oh \etal\ 2008}\nocite{oh08} and application in {Weisz
  \etal\ 2009b}\nocite{weisz09b}).  After removing obvious foreground
stars, the 3.6 $\mu$m image (which also shows the truncation of the
stellar population in the southern region of the system; see Figure~4
of {Walter \etal\ 2007}\nocite{walter07}) was spatially smoothed to
7.5\arcsec\ resolution (our highest \HI\ angular resolution element)
and then fitted with ellipses of constant surface brightness.  The
best-fit solution yielded a central position of ($\alpha$,$\delta$) =
(13:06:24.7, 67:42:33.2), an ellipticity of 0.5, and a position angle
of 90$^{\circ}$.  The radially averaged 3.6 $\mu$m surface brightness
profile is shown in Figure~\ref{figcap17}; the surface brightness
declines with radius, falling from $\sim$21 mag\,arcsec$^{-2}$ to
$\sim$23 mag\,arcsec$^{-2}$ within $\sim$3 kpc.

The \HI\ scale height of an isothermal disk is related to the velocity
dispersion of the gas and the mass surface density.  As derived in
\citet{kellman70} and applied in \citet{kim99} and \citet{weisz09b},

\begin{equation}
h=\frac{<\sigma_{v}^2>}{\pi {\rm G} \Sigma},
\label{eq3}
\end{equation}

\noindent where $\sigma_{\rm V}$ is the \HI\ velocity dispersion in
\kms, $\Sigma$ is the mass density profile in \msun\,pc$^{-2}$, and 
{\it h} is the \HI\ disk scale height in pc.  Using the same central 
position, ellipticity and position angle used
to derive the stellar surface brightness profile, we extracted
radially averaged values of \HI\ column density and velocity
dispersion.  These plots are shown in Figure~\ref{figcap17}.  
The rise in \HI\ velocity dispersion in the central $\sim$0.5
kpc, and the prominent low column densities in the same region, 
are the results of the giant \HI\ hole. We note that the observed
velocity dispersions may not be appropriate for calculating the 
scale height of the gas prior to the onset of SF; we thus also perform
energy calculations with assumed velocity dispersions similar to those
seen in the quiescent regions of nearby dwarf galaxies (see further 
detailed discussion below).

\begin{figure}
\epsscale{1.0}
\plotone{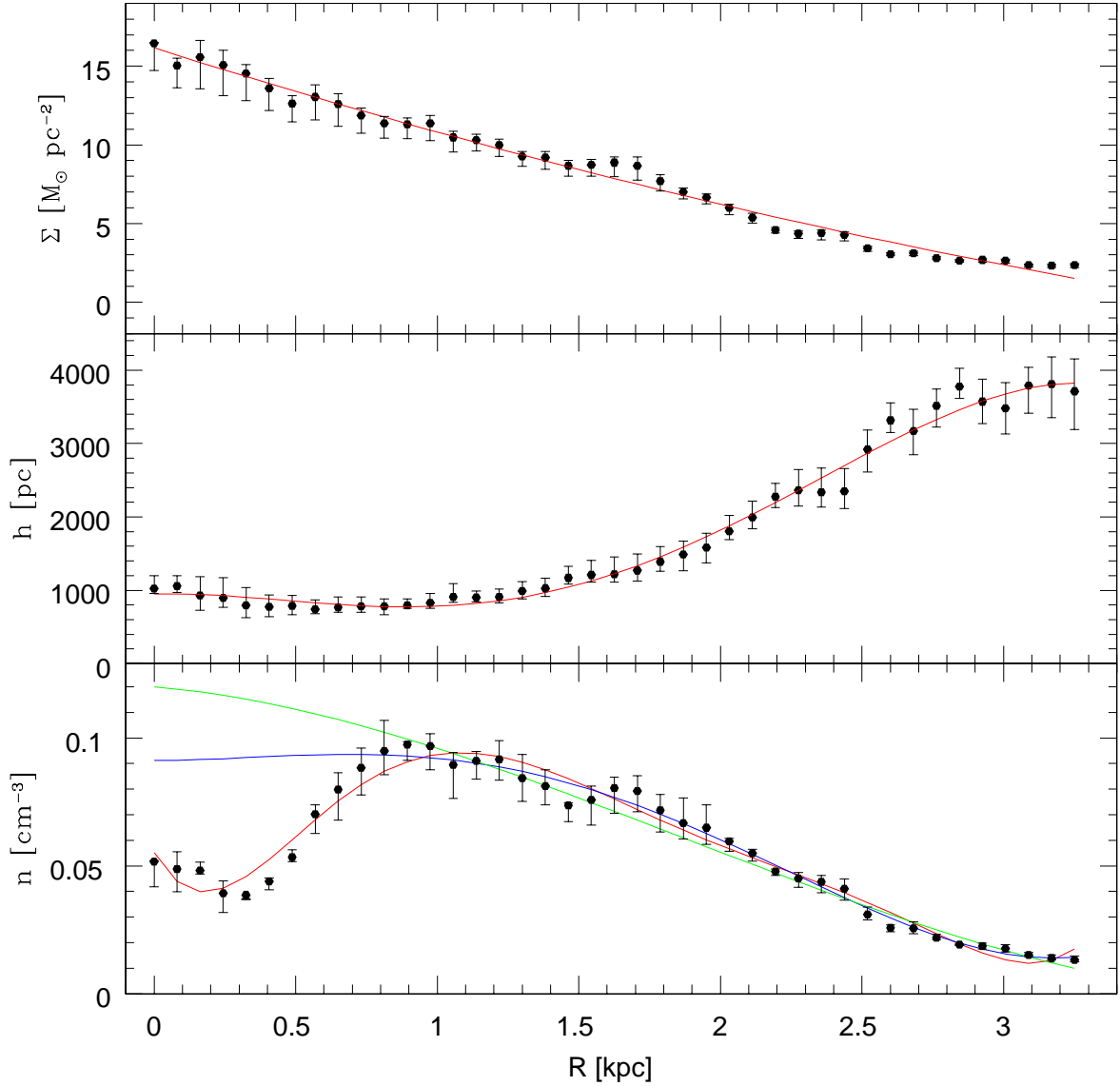}
\epsscale{1.0}
\caption{{\it Top}: Radial mass surface density profile derived from
  the 3.6 $\mu$m surface brightness profile (Figure~\ref{figcap17}),
  using the method of \citet{oh08}.  {\it Middle}: Radial scale height
  profile derived from the mass surface density and the velocity
  dispersion profiles (Figure~\ref{figcap17}) assuming an isothermal
  disk.  This derivation used the method of Kellman (1970; see also
  Kim \etal\ 1999).  {\it Bottom}: Radial \HI\ volume density ($n_0$)
  profile.  The points on all of the graphs are the derived values and
  the red curves are functional fits to the data.  The function shown in
  blue is fitted to the data beyond 1 kpc and is forced to plateau to
  a constant volume density in the inner disk; the function in green
  is also fitted to the data beyond 1 kpc using a Gaussian function
  that rises to slightly larger n$_{\rm 0}$ values in the inner disk
  (see discussion in \S~\ref{S4.2}.}
\label{figcap18}
\end{figure}

Figure~\ref{figcap18} shows radial profiles of derived properties
derived from observed properties shown in Figure~\ref{figcap17} and
using the equations above.  The top panel of Figure~\ref{figcap18}
shows the mass surface density in units of \msun\,pc$^{-2}$, derived
from the 3.6 $\mu$m surface brightness profile by assuming a 3.6
$\mu$m mass to light ratio of 0.5 (similar to other dIrr systems; see
{de~Blok \etal\ 2008}\nocite{deblok08}) and applying the techniques
used in \citet{oh08} and \citet{weisz09b}.  The middle panel of
Figure~\ref{figcap18} shows the \HI\ scale height, {\it h}, as a
function of radius, derived using the radial values of $\sigma_{\rm
  V}$ and $\Sigma$.  The shape of the \HI\ scale height as a function
of radius is qualitatively similar to the profile derived for the
M\,81 group dwarf Holmberg\,II in \citet{weisz09b} using a similar
approach; however, the scale height in the outer disk of DDO\,165 is a
factor of $\sim$3 larger than for Holmberg\,II.  The red line shows
the polynomial fit to this curve, which we use to obtain the value of
{\it h} at any given radius.

Finally, the bottom panel of Figure~\ref{figcap18} shows the midplane
\HI\ volume density as a function of radius, derived using the radial
values of N$_{\rm HI}$ and {\it h}. The giant \HI\ hole is prominent
at radii $<$1 kpc.  Moving outward toward r $=$ 1 kpc, the scale height
remains essentially constant, $\Sigma$ is decreasing, $\sigma_{\rm V}$
falls to a more or less constant value, and the \HI\ column density
rises.  These properties conspire to create a maximum in the
\HI\ volume density at r $\sim$ 1 kpc.  The volume density then falls
smoothly moving further outward through the disk; the increasing scale
height implies lower \HI\ volume densities at these large radii.

Three fits to the radial plot of \HI\ volume density are shown in
Figure~\ref{figcap18}.  The red line shows the functional fit including
all data points.  Since there was \HI\ gas inside this structure when
the BHeB stars were formed, our derivation of the energy
requirements for the formation of this hole would be an underestimate
if we used the present-day volume density.  Thus, we include two other
fits to the n$_{\rm 0}$ curve: the function shown in blue is fitted to
the data beyond 1 kpc and is forced to plateau to a constant volume
density in the inner disk; the function in green is also fitted to the
data beyond 1 kpc using a Gaussian function that rises to slightly
larger n$_{\rm 0}$ values in the inner disk.  While these fits do not
explicitly conserve \HI\ mass, we stress that the volume density
enters the energy calculation as n$_{\rm 0}^{1.12}$; the choice
between these three curves produces significant differences in energy
only in the inner regions of the galaxy (i.e., where the red and
blue/green curves differ substantially).

The radial behavior of \HI\ volume density now allows us to estimate
the creation energies for the \HI\ holes.  We begin by considering
those 3 holes with measured expansion velocities (2, 5, 7).  Knowing
these velocities and the sizes of the holes (radius for circular
structures, geometric radius for elliptical structures), we simply
insert the volume density values at the appropriate galactocentric
radius (Figure~\ref{figcap18}) into Equation~\ref{eq1}.  The results
are shown in Table~\ref{t4}; for each hole, we estimate the requisite
creation energy assuming four values of the \HI\ volume density.  The
simplistic constant n$_{\rm 0}$ approach (as in Paper I) is shown in
columns 3 and 4; the polynomial fit approach is shown in columns 5 and
6; the best fit line at r $>$ 1 kpc and plateau to constant n$_{\rm
  0}$ at r = 0 approach is shown in columns 7 and 8; the best fit
Gaussian profile at r $>$ 1 kpc approach is shown in columns 9 and 10.

Three important insights arise from an examination of Table~\ref{t4}
and Figure~\ref{figcap18}.  The first is that the most significant
differences in volume density occur between the simplistic n$_{\rm 0}$
= 0.1 cm$^{-2}$ assumption (close to the value at r $\simeq$ 1 kpc)
and the outer disk regions.  This implies that estimates of the
energy requirements for a hole in the outer disk will be
significantly lower using the mass model approach discussed above than
using the assumed value from Paper I.  Hole 7 shows this effect
quite clearly, as the requisite energies differ by a factor of two.
The second insight is that for all holes at radii beyond $\sim$1 kpc,
the functional fits in Figure~\ref{figcap18} use the same data and thus
produce essentially identical n$_{\rm 0}$ values and energy
requirements.  The final important feature of Table~\ref{t4} is that
the creation of each of these kinematically expanding structures
requires of order 10$^{51}$ erg.  This estimate is similar to the
energy requirements for the smaller holes in Holmberg\,II found by
\citet{weisz09b}.  

We next consider those holes that show no signs of kinematic
expansion, rendering a direct application of Equation~\ref{eq1}
difficult.  These kinematically static structures are usually assumed
to have either blown out of the disk or to have re-established
pressure equilibrium with the surrounding ISM.  The latter effect will
occur when the average bulk velocity of a shell equals the average
velocity dispersion over a given region.  Thus, if we assume coherent
motion at an average velocity equal to that velocity dispersion, we
can estimate the energy requirements for the structure.  We stress
that such estimates are uncertain at the 50\% level or more;
nonetheless these numbers are useful when compared to the mechanical
energy injected by the evolving massive star populations (see below).

In an attempt to bracket the potential extrema in the assumed velocity
ranges, we consider three different values for the putative expansion
velocities of the kinematically stalled structures.  First, we assume
a value of 7 \kms\ for any stalled hole, as adopted by the analyses in
The HI Nearby Galaxies Survey (THINGS; {Walter
  \etal\ 2008}\nocite{walter08}).  Next, we can adopt one of two
different values for the velocity dispersion: $\sigma_{\rm V}$
averaged over the entire hole structure, or $\sigma_{\rm V}$ averaged
over the entire galaxy.  As expected, the global $\sigma_{\rm V}$
value (8.5\,\pom\,5.0 \kms, regardless of angular resolution) is lower
than in localized regions ($\sigma_{\rm V}$ $\simeq$ 11-12 \kms).

The energy estimates for the stalled structures are summarized in
Table~\ref{t5}.  Column 2 gives the assumed expansion velocity for
each hole; the remaining columns contain the same information as those
in Table~\ref{t4}: constant n$_{\rm 0}$ in columns 3 and 4; the
best-fit polynomial function in columns 5 and 6; the best fit function
plateaus to constant n$_{\rm 0}$ in columns 7 and 8; the best fit
Gaussian in columns 9 and 10.  The table repeats these calculations
for each assumed V$_{\rm exp}$ value.  Similar trends are seen for
these stalled holes as described above: the largest differences at a
chosen velocity occur between holes in the outer disk and holes at the
maximum n$_{\rm 0}$ value; energy estimates for holes in the outer
disk are independent of the adopted n$_{\rm 0}$ value.  Further, for
the smaller \HI\ holes in Table~\ref{t5} (1, 4) the estimated
creation energies are similar to those found for the expanding
features in Table~\ref{t4}.  Note by comparing the energy estimates
for a given hole that the V$_{\rm exp}^{1.4}$ dependence results in a
factor of $\sim$2 variance in the derived energy budget.

The giant \HI\ hole (Hole 6) stands out prominently in Table~\ref{t5}.
As expected, the characteristic creation energy of this structure is
roughly two orders of magnitude larger than the other holes.  Further,
this characteristic energy of $\sim$10$^{53}$ erg is similar to those
derived for other giant \HI\ structures in dwarf galaxies (e.g.,
{Walter \& Brinks 1999}\nocite{walter99}; {Weisz
  \etal\ 2009b}\nocite{weisz09b}; {Warren
  \etal\ 2011}\nocite{warren11}).  

The information in Tables~\ref{t4} and \ref{t5} allows us to estimate
kinematic ages for each of these structures; these ages are given in
Table~\ref{t6}.  For the expanding holes (Table~\ref{t4}) this number
is simply the geometrical mean radius divided by the measured
expansion velocity (assumed to be constant).  Note that these ages are
explicitly upper limits since the expansion velocity was higher
immediately after the hole was created.  For the stalled structures,
we adopt the global velocity dispersion as being representative of the
current expansion velocity and use it in subsequent discussion.  This
is motivated by two considerations.  First, the global $\sigma_{\rm
  V}$ value provides energy estimates near the middle of the ranges
shown in Table~\ref{t5}.  Second, the galaxy-averaged velocity
dispersion sets the characteristic ISM pressure and thus determines in
large part when an expanding \HI\ structure will stall.  The kinematic
ages tabulated in Table~\ref{t6} range from 22 to 89 Myr.  This time
interval corresponds to the period of heightened SFR between 25-100
Myr discussed above; it also sets the characteristic timescale of 100
Myr over which we will compare the energy outputs from stellar
evolution with the energy requirements from the \HI\ data.

In the discussion that follows we adopt the energy estimates
derived using the Gaussian profile for \HI\ volume density of all
\HI\ holes.  For expanding holes, V$_{\rm exp}$ is an observed
quantity.  As justified above for the stalled holes, we hereafter set
V$_{\rm exp}$ $=$ 8.5 \kms, i.e., equivalent to $\sigma_{\rm V}$ for
the entire galaxy.  These assumptions lead to the E$_{\rm Hole}$
values in column 3 of Table~\ref{t6}.  While the relative errors on
these values are difficult to estimate, we stress that our
calculations should be considered uncertain at the 50\% level (at
minimum); the scatter in E$_{\rm Hole}$ estimates from Tables~\ref{t4}
and \ref{t5} can also be used to guide interpretation.  We emphasize
that the specific E$_{\rm Hole}$ values are less important than the
order of magnitude estimates; \S~\ref{S4.3} shows that the feedback
energies from the stellar populations inside each hole exceed these
E$_{\rm Hole}$ estimates, usually by more than an order of magnitude.

\subsection{The Energies of Recent Star Formation}
\label{S4.3}

The global and localized SFHs presented above allow us to calculate
the integrated mass of the stars formed as a function of time; this is
simply the product of average SFR and time bin width.  With the
assumption of a fully populated universal Salpeter IMF, this total
stellar mass implies a certain number of massive stars that will
result in SNe and therefore deposit mechanical energy into the
surrounding ISM.  The amount of feedback energy resulting from the
evolution of these massive stars can thus be quantified using our SFHs
and models of the energy outputs of SF episodes.  To this end, we
apply the STARBURST99 models \citep{leitherer99} to each of the SFHs
described in \S~\ref{S3} above.  Specifically, we perform a simulation
of a 10$^6$ \msun\ instantaneous burst of SF, scaled to the integrated
mass in each SFH, using the same relevant parameters as discussed in
\S~\ref{S3.3}.  It is assumed that each star more massive than 8
\msun\ explodes as a SN \citep{salaris05,smartt09}.  We make no
correction for the contributions of Type Ia SNe to the feedback energy
budget.  These events occur over a range of timescales
\citep{matteucci06} and may occur in regions that are already
partially evacuated; for both reasons, the actual feedback energy from
stellar evolution in a given region can be considered to be a lower
limit.  The STARBURST99 models are sampled every 5 Myr in order to
create profiles of the feedback energy resulting from SF as a function of time.

Our {\it HST} imaging provides detailed information on the energy
budget as a function of time; however, we are cognizant of the
uncertainties introduced as a result of random stellar diffusion.  As
stars move within a galaxy, they will disperse away from their
formation sites; this effect becomes important as stars venture into
and out of given areas of consideration within a galaxy. While the
results presented by \citet{bastian11} suggest a long period of
coherence of stellar structure within DDO\,165 (see discussion above),
we conservatively limit our consideration of the input energies
within a given region to 100 Myr.  By considering only this recent
interval, we can minimize the uncertainties resulting from stars
diffusing into or out of a given region.  By way of example, the
control fields have physical radii of $\sim$220 pc at the adopted
distance.  A 2 km\,s$^{-1}$ net velocity will move this distance in
$\sim$100 Myr; however, the relaxation dynamics are likely complex and
the actual crossing time will depend on many paramaters (local stellar
and gas density, burst mass and age, etc.).  For the larger \HI\ holes
we observe in DDO\,165 (especially \#6, with a minor axis of $\sim$1.1
kpc), the effects of diffusion out of the region will be smaller yet.

\begin{figure}
\epsscale{1.0}
\plotone{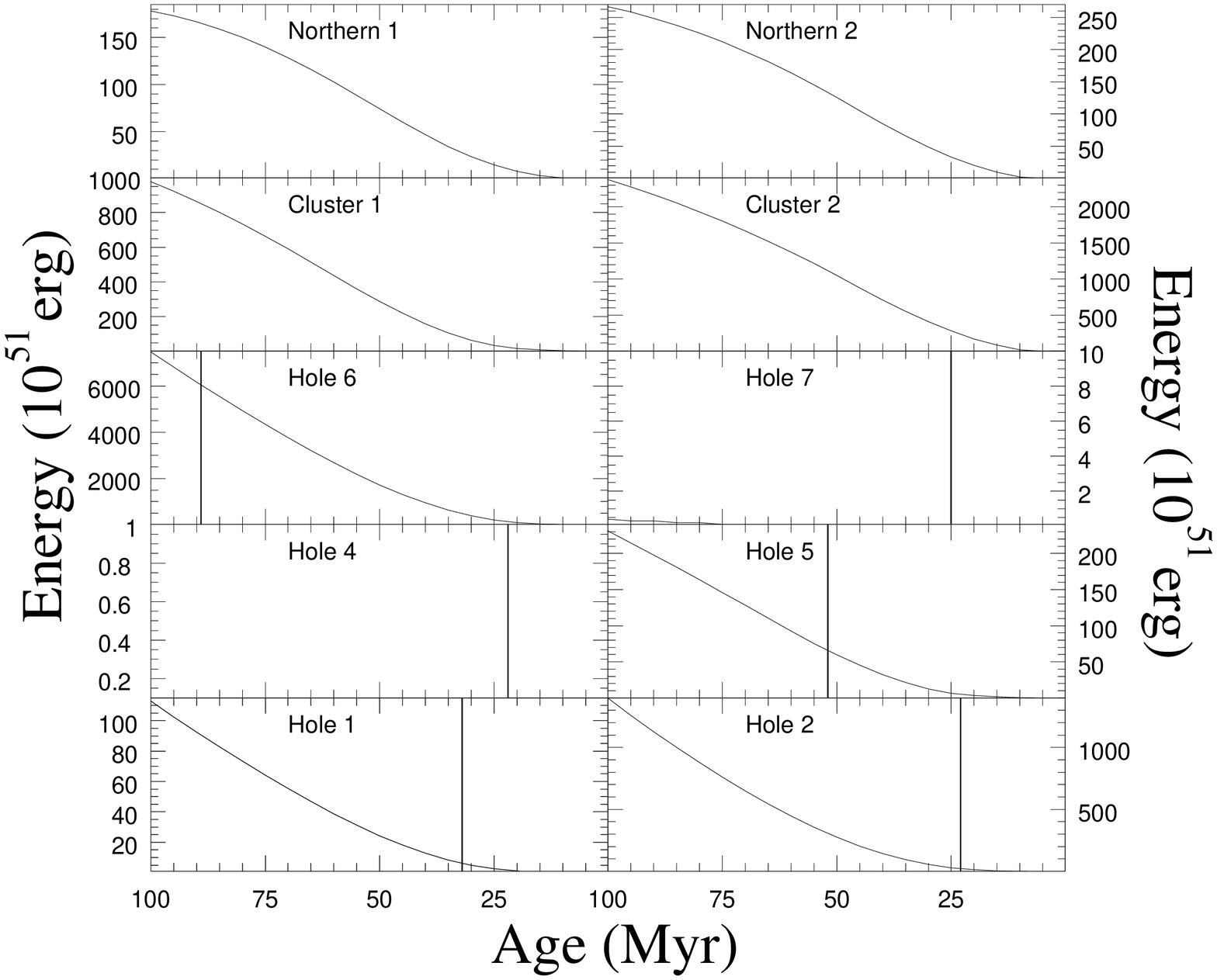}
\epsscale{1.0}
\caption{Cumulative feedback energy profiles of localized regions, as
labeled; these profiles are computed using STARBURST99 (Leitherer
\etal\ 1999) with the measured SFHs as inputs; see detailed discussion
in \S~\ref{S4.3}. The energies are integrated from the present (t = 0)
to lookback times of 100 Myr.  The CMDs and SFHs of these fields are 
shown in Figures~\ref{figcap5} and \ref{figcap9}, respectively.}
\label{figcap19}
\end{figure}

\begin{figure}
\epsscale{1.0}
\plotone{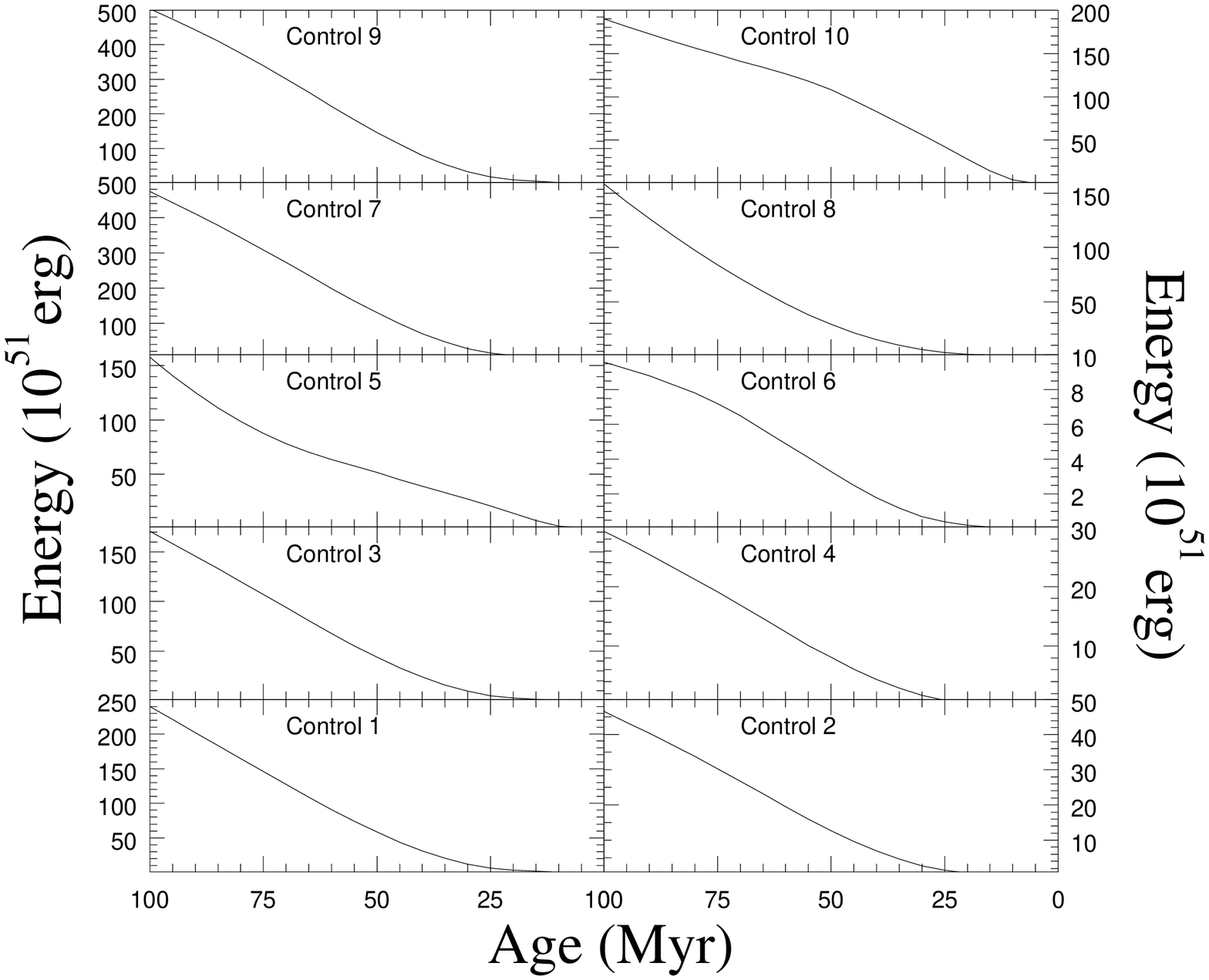}
\epsscale{1.0}
\caption{Cumulative feedback energy profiles of localized regions, as
labeled; these profiles are computed using STARBURST99 (Leitherer
\etal\ 1999) with the measured SFHs as inputs; see detailed discussion
in \S~\ref{S4.3}. The energies are integrated from the present (t = 0)
to lookback times of 100 Myr.  The CMDs and SFHs of these fields are 
shown in Figures~\ref{figcap6} and \ref{figcap10}, respectively.}
\label{figcap20}
\end{figure}

\begin{figure}
\epsscale{1.0}
\plotone{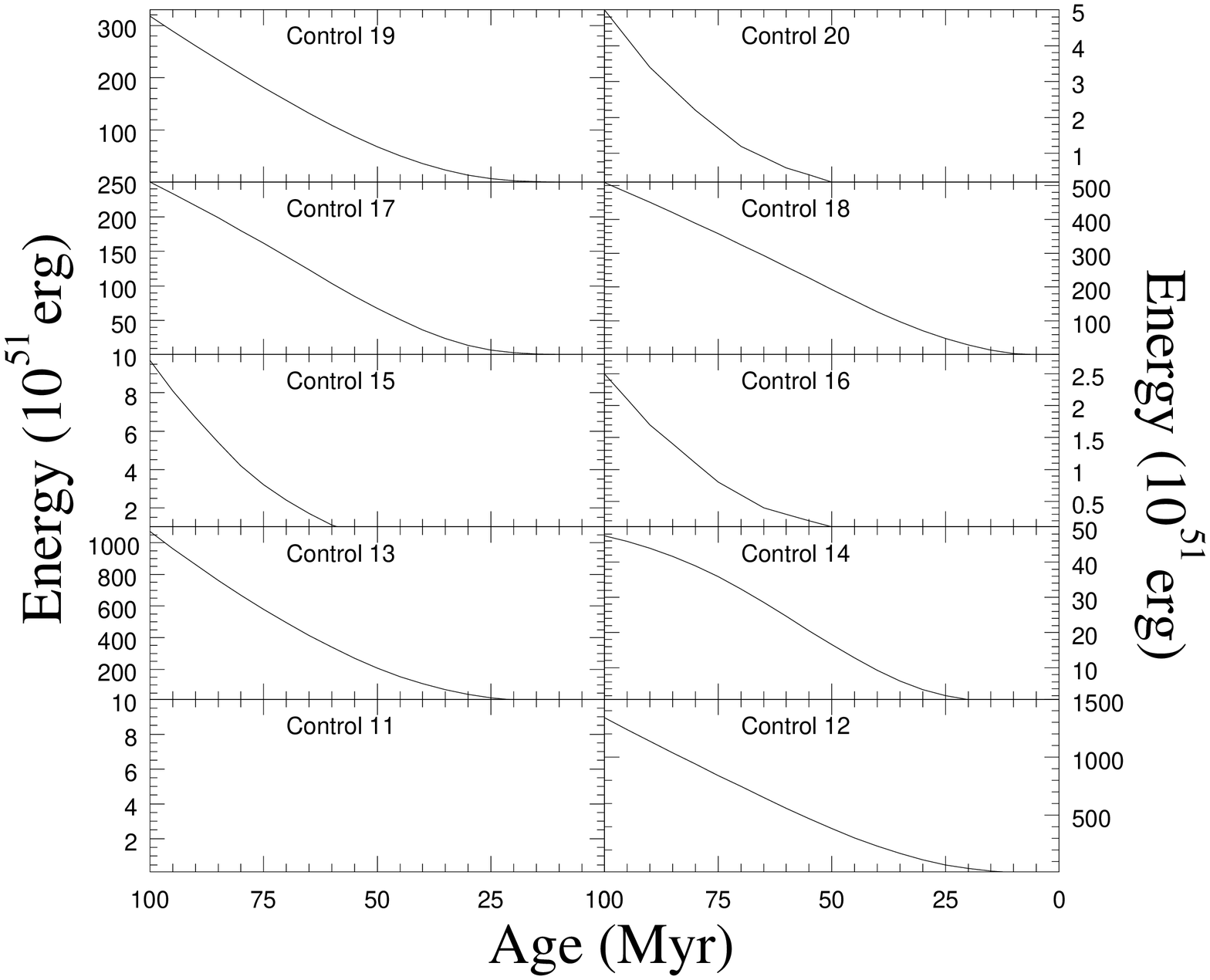}
\epsscale{1.0}
\caption{Cumulative feedback energy profiles of localized regions, as
labeled; these profiles are computed using STARBURST99 (Leitherer
\etal\ 1999) with the measured SFHs as inputs; see detailed discussion
in \S~\ref{S4.3}. The energies are integrated from the present (t = 0)
to lookback times of 100 Myr.  The CMDs and SFHs of these fields are 
shown in Figures~\ref{figcap7} and \ref{figcap11}, respectively.}
\label{figcap21}
\end{figure}

The results of our energy calculations over the most recent 100 Myr
are shown in Figures Figures~\ref{figcap19}, \ref{figcap20}, and
\ref{figcap21}.  A comparison of these profiles with the normalized
SFH plots (Figures~\ref{figcap9} through \ref{figcap11}) demonstrates
several interesting points.  First, total feedback energies over the
last 100 Myr fall in the $\sim$10$^{51}$--10$^{54}$ erg range.
Further, the energy input into each of these regions is temporally
extended; the canonical assumption of an instantaneous injection of
feedback energy during the formation of a stellar cluster is a
simplification compared to the extended energy profiles.  Note that
these profiles could be extended to larger lookback times (with
appropriate caveats), increasing the total energy feedback budgets.

The cumulative energy profiles shown in Figures~\ref{figcap19} through
\ref{figcap21} allow us to quantify the amount of energy resulting
from recent SF on any timescale of interest between 0 and 100 Myr.
These energies can then be directly compared with the energy
requirements for creating the \HI\ structures (see discussion in
\S~\ref{S4.2} above).  Recall that Table~\ref{t6} shows the adopted
kinematic ages of the \HI\ structures (column 2), as well as the
requisite energies to create them (column 3).  It is now
straightforward to find the amount of energy produced by SF and
stellar evolution over the adopted kinematic ages of the structures:
one simply reads the value of energy at the appropriate age in the
panels of Figure~\ref{figcap19}.  To guide interpretation, that figure
shows the kinematic age of the structure by a vertical line.  Column 4
of Table~\ref{t6} tabulates E$_{\rm SF}$, the total energy input by
stellar evolution over the adopted kinematic age of the
\HI\ structure.

Note by examining the SFHs and energy profiles of \HI\ Holes 4
(stalled structure) and 7 (expanding structure) that an estimate of
E$_{\rm SF}$ is not possible for these features: no recent SF has
occurred within the kinematic ages of these holes.  This might not be
surprising for Hole 4: it is located in a region of low \HI\ column
density and low stellar density.  Hole 7 is much more difficult to
interpret; it contains dense neutral gas (overlapping the 10$^{21}$
cm$^{-2}$ contour) and is expanding at 9 \kms.  However, it hosts only
low-level SF over the past 500 Myr.

For the other four \HI\ structures, the energy derived from SF over
the kinematic age is larger than the energy required to create the
structure.  We term the ratio E$_{\rm Hole}$/E$_{\rm SF}$ the
``feedback efficiency''.  The efficiency is relatively high for hole 1
($\sim$40\%) but low for the giant \HI\ hole ($\sim$1\%); these
efficiencies are similar to those found for holes in Holmberg\,II by
\citet{weisz09b}.  For these \HI\ structures, recent stellar evolution
provides sufficient energies within the kinematic ages of the hole in
order to create them.  We note that E$_{\rm SF}$ $>$ E$_{\rm Hole}$
for any of the energy estimates in Tables~\ref{t4} or \ref{t5},
even when selecting the highest E$_{\rm Hole}$ values.  The assumption
of 100\% efficiency is implicit; smaller values of efficiency produce
smaller energy excesses, but retain sufficient cumulative energy
inputs to create the observed structures.

As noted above, we can only estimate the kinematic ages of structures
that are expanding at the present time.  As such, a second timescale
of interest is the amount of time required for the energy from SF to
equal the energy required to create the structure.  We hereafter refer
to this characteristic timescale as t$_{\rm eq}$, the time when
E$_{\rm SF}$ $=$ E$_{\rm Hole}$ (assuming 100\% efficiency); we
tabulate the values of t$_{\rm eq}$ in column 6 of Table~\ref{t6}.
This time interval is very brief for those structures with SF in the
last $\sim$50 Myr: the time required is less than 30 Myr in all cases.
For those structures without recent SF (Holes 4 and 7), this interval
is much longer: $\sim$130 Myr for Hole 7 and $\sim$330 Myr for Hole
4. These extended intervals should be treated with caution; as noted
above, on timescales longer than 100 Myr, random stellar motions of a
few km\,s$^{-1}$ can move stars both into and out of areas of the size
we consider here.

We can envision a second and related source of uncertainty in the
interpretation of the energy timescales.  Consider a concentrated
region of SF that imparts energy into the surrounding ISM and creates
an expanding structure; as the hole diameter increases, it can cover a
much larger area than the SF that created it. If there were any recent
SF in this surrounding area, then the energy budget increases,
although that energy did not specifically contribute to the creation
of the \HI\ hole.  In this sense the energy estimates of our
structures may be overestimates.  However, it seems likely that large
contributions from concentrated SF would create \HI\ structures of
their own; as the two structures merge they create a single structure.
Further, we make no corrections for the energy contributions from
Type Ia SNe (see discussion above), arguing for the derived energies
being underestimated.  With these uncertainties in mind
for Holes 4 and 7, we note that t$_{\rm eq}$ is insensitive to the
adopted E$_{\rm Hole}$ value: again selecting the largest E$_{\rm
  Hole}$ values from Tables~\ref{t4} and \ref{t5} only changes t$_{\rm
  eq}$ for Holes 1 (t$_{\rm eq}$ $<$ 35 Myr), 4 (t$_{\rm eq}$ $<$ 400
Myr), and 6 (t$_{\rm eq}$ $<$ 25 Myr).

A direct comparison of the energies from SF with the energies of
the surrounding gas is by definition limited to the \HI\ holes only.
However, following the discussion above, we can consider the feedback
energies injected over only the most recent $\sim$100 Myr in any
region within DDO\,165.  We thus tabulate in Table~\ref{t7} the
feedback energies from SF after 25, 50 and 100 Myr in each of the
regions shown in Figure~\ref{figcap4}.  This allows a comparison of
the energies from recent SF in different regions throughout the
galaxy in a straightforward manner.

Multiple interesting trends are seen in the data presented in
Table~\ref{t7}.  First, most regions have accumulated of order
10$^{52}$ erg of energy by 25 Myr and of order 10$^{53}$ erg of energy
by 100 Myr.  This enforces the importance of temporally extended SF on
the surrounding ISM \citep[e.g.,][]{mcquinn10}.  Second, the regions
with ongoing SF within the most recent intervals stand out
prominently: for example, Cluster\,2 has the largest E$_{\rm SF}$
value at 25 Myr (2.8\,$\times$\,10$^{53}$ erg), exceeding even the
feedback energy within the much larger giant \HI\ hole.  Remarkably,
this substantial energy budget has failed to produce a coherent
\HI\ structure at our spatial and velocity resolution (see further
discussion below).  Third, the increase in the cumulative E$_{\rm SF}$
values between 25 and 100 Myr (when E$_{\rm SF}$ at 25 Myr is not
zero) is at least an order of magnitude for many (but not all)
regions. This again suggests that the stellar-ISM interaction operates
over timescales much longer than the lifetimes of massive stars.
Finally, and perhaps most importantly, the control fields have similar
energy budgets on these timescales as do the \HI\ holes.  Taken at
face value, the data in Table~\ref{t7} suggest that abundant feedback
energy is injected throughout the disk of DDO\,165; coherent
\HI\ structures form in some regions but not others.

\section{Star Formation As An Evolutionary Catalyst}
\label{S5}

The above discussion has demonstrated that significant feedback energy
is produced by stellar evolution processes in regions throughout
DDO\,165 that span a large range of physical parameter space.  In the
discussion that follows, we discuss the stellar feedback hypothesis as
applied to both the \HI\ holes and the regions located throughout the
system.  While alternatives to the feedback hypothesis do exist (e.g.,
high velocity cloud impacts, disk instabilities, turbulence, ram
pressure stripping; {S{\'a}nchez-Salcedo
  2002}\nocite{sanchezsalcedo02} and references therein), our data are
optimal for examining the energies of stellar evolution and of
\HI\ structures.  Multiple recent works have applied similar
techniques in studying the feedback mechanism in nearby galaxies
\citep[e.g.,][]{weisz09b,bagetakos11}; we refer the reader to the
references in those works (as well as in Paper I) concerning potential
alternative creation mechanisms.

\subsection{Regions With Coherent \HI\ Structure}
\label{S5.1}

\subsubsection{\HI\ Holes}
\label{S5.1.1}

We now address a deceivingly simple question: are the observed
\HI\ holes the results of SF that occurred in the past 100 Myr?  We
consider each structure individually to address this question.  Three
holes are stalled (1, 4, 6) and three are expanding (2, 5, 7).

Hole 2 can be considered the clearest example of feedback energy
creating an \HI\ hole in DDO\,165.  The high stellar density and
dramatic increase in SFR during the 100-25 Myr interval produce a
rapid deposit of mechanical energy into the surrounding ISM.  The
result is the most rapidly expanding structure identified in this
study.

Holes 4 and 7 have no SF within the most recent $\sim$100 Myr, and
thus represent challenges to the hypothesis of creation via SF
feedback.  Sufficient stellar feedback energies are only injected over
longer timescales (t$_{\rm eq}$ $\simeq$ 330 Myr for Hole 4 and
$\simeq$130 Myr for Hole 7; see Table~\ref{t6}).  If these structures
can remain coherent over timescales of hundreds of Myr (e.g., {Recchi
  \& Hensler 2006}\nocite{recchi06}), then the creation of these
structures by stellar evolution processes poses no apparent
difficulties. The location of Hole 4 in the outer disk and the low
surrounding \HI\ surface densities might argue in favor of such a
long-lived coherent structure. However, the location of Hole 7 poses a
significant issue: given the global velocity dispersion value
($\sigma_{\rm V}$ $=$ 8.5 \kms), the crossing time of that structure
is of order 25 Myr.  This implies that the surrounding gas will
re-establish pressure equilibrium on similar timescales.  Further, the
expanding nature of the structure is challenging to reconcile with the
lack of recent SF over the most recent $\sim$100 Myr.  We interpret
these structures as evidence against a recent ($\lsim$100 Myr)
feedback origin; however, feedback on longer timescales remains a
viable mechanism.

Holes 1 and 5 present a similar juxtaposition; 5 is expanding while 1
is stalled.  Both structures have E$_{\rm SF}$ $>$ E$_{\rm Hole}$,
implying that SF is energetically capable of producing the holes.
Hole 1 has a higher normalized SFR level and a higher stellar density
than Hole 5.  Hole 5, on the other hand, has experienced a recent
period of slightly elevated SFR (100-25 Myr); this can be interpreted
as a potential link between recent SF and hole creation.

The giant \HI\ hole is perhaps the most challenging structure to
interpret in DDO\,165.  The energies derived from SF are clearly
energetically capable of removing the \HI\ gas from within.  However,
the size of the structure, the low column densities within, and the
presence of recent SF cloud this simplistic interpretation.  In order
for stars to have formed recently, the structure must either have
retained interior local \HI\ overdensities beyond our resolution limit
or have expanded to its present size extremely rapidly (within
$\sim$50 Myr, implying V$_{\rm exp}$ $\simeq$25 \kms, easily
detectable at our velocity resolution).  Alternatively, it is possible
that the elevated SFR in the 400-300 Myr interval caused the ISM to
become highly fragmented (such as in IC\,2574 and Ho\,II; see {Weisz
  \etal\ 2009a}\nocite{weisz09a}, {2009b}\nocite{weisz09b}); the
feedback energy injected by the more recent SF then fully evacuated
the remaining gas.

Taken as a group, we conclude that the coherent \HI\ holes identified
in Paper I can be created by stellar evolution processes, so long as
the structures can remain coherent over 100 Myr timescales.
Stated equivalently, our data are in agreement with a model where a
mixed age and spatially distributed stellar population produces SNe
separated by tens or hundreds of Myr.  The cumulative energy
outputs of these populations of stars are responsible for the creation
of the \HI\ holes.  This is in agreement with the conclusions of
\citet{weisz09b} based on a larger sample of holes in the M\,81 group
dwarf Holmberg\,II, as well as with the conclusions of
\citet{warren11} for a collection of dIrr galaxies with giant
\HI\ holes.

\subsubsection{The Kinematically Distinct Northern Region}
\label{S5.1.2}

We now consider the origin of the kinematic discontinuity between the
northern and southern \HI\ components.  As Paper I shows, these
components are static with respect to one another along the line of
sight.  Is this separation the result of macroscopic motion of
\HI\ gas due to SF (i.e., outflow) or is it the pristine infall of an
\HI\ cloud?  Paper I uses arguments based on the \HI\ data alone to
eliminate various potential origins for the northern cloud (including
superposed or counter-rotating disks, a coherently rotating disk, and
turbulence).  However, based on the \HI\ data alone, we cannot
differentiate between infall and outflow origins for the northern
component.

\begin{figure}
\epsscale{1.0}
\plotone{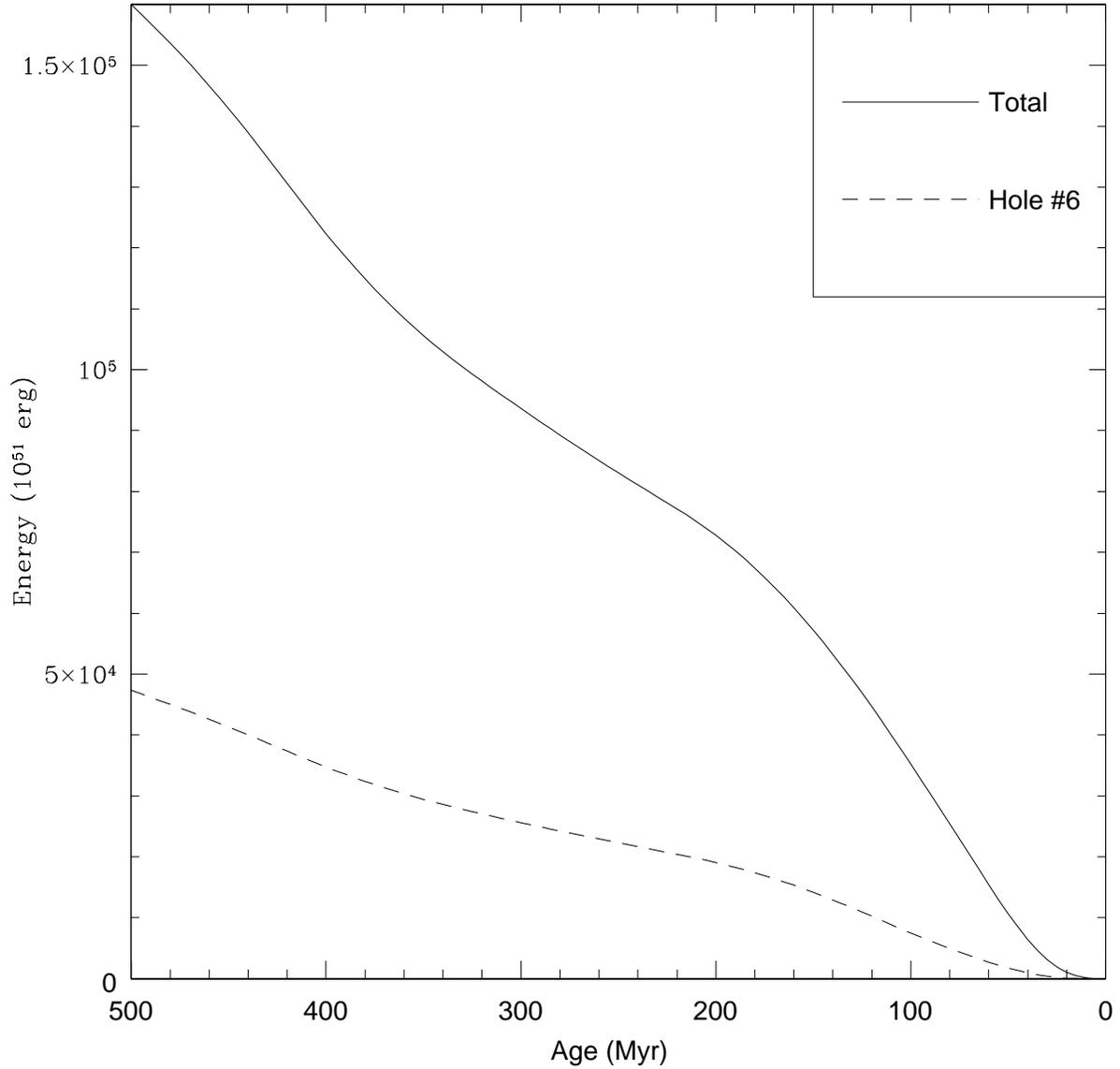}
\epsscale{1.0}
\caption{Cumulative feedback energy profiles of the entire galaxy
(solid line) and \HI\ Hole 6 (dashed line); these profiles are
computed using STARBURST99 (Leitherer \etal\ 1999) with the measured
SFHs as inputs; see detailed discussion in \S~\ref{S4.3}. The energies
are integrated from the present (t = 0) to lookback times of 500 Myr.}
\label{figcap22}
\end{figure}

Our energy analysis based on the recent SFH provides tentative
evidence for the outflow scenario.  As shown in Figure~\ref{figcap22},
within the last 500 Myr (roughly half of the burst duration; see
{McQuinn \etal\ 2010}\nocite{mcquinn10}), the stars within DDO\,165
have injected $\sim$10$^{56}$ erg of energy into the ISM.  This
cumulative energy budget is in principle capable of moving the
$\sim$1.65\,$\times$\,10$^{7}$ \msun\ associated with the northern
component (see Paper I) $\sim$1 kpc out of the total gravitational
well of DDO\,165.  We estimate the required energy via a simple
gravitational potential energy calculation:

\begin{equation}
{\rm E = {(8.56\,\times\,10^{37})}{\cdot}\frac{M_{north}{\cdot}M_{dyn}}{R}\,\,\,(erg)}
\label{eq4}
\end{equation}

\noindent where 8.56\,$\times$\,10$^{37}$ is a constant that
incorporates G and conversion factors, M$_{\rm north}$ is the mass of
the northern \HI\ component in \msun, M$_{\rm dyn}$ is the total mass
of the DDO\,165 system in \msun, and R is the separation between the
northern \HI\ component and the center of the gravitational potential
well.  Assuming M$_{\rm north} =$ 1.65\,$\times$\,10$^7$ \msun,
M$_{\rm dyn} =$ 10$^9$ \msun\ (i.e., $\sim$10 times larger than the
total \HI\ mass), and R $=$ 1 kpc, this yields E $\simeq$ 10$^{56}$
erg.  We stress, however, that this calculation requires the
assumption of various parameters that are not observationally
constrained; some of the most important will be total depth and shape
of the gravitational potential, and potential relative motion of the
northern and southern components out of our line of sight (recall that
the inclination of the system remains unconstrained) or at projected
velocities lower than our spectral resolution limit.  Our stellar
energies thus favor the outflow hypothesis, but we treat this
conclusion with caution.

The outflow interpretation is strengthened considerably by the nebular
spectroscopy presented in \citet{croxall09}.  There, two slit
positions targeted the highest surface brightness \halpha\ regions;
these positions ([13:06:16.42, $+$67:42:08.4] and [13:06:24.52,
  $+$67:43:41.2]) correspond to Cluster 2 in the main body and Northern
1/Northern 2, respectively.  The derived oxygen abundances of these
two regions are identical within the errors (12 $+$ log(O/H) $=$
7.79\,\pom\,0.20 and 7.81\,\pom\,0.20).  This supports the outflow
interpretation based on the well-known chemical uniformity seen in
most dwarf galaxies (see {Kobulnicky \& Skillman
  1996}\nocite{kobulnicky96}, {1997}\nocite{kobulnicky97b},
{Kobulnicky \etal\ 1997}\nocite{kobulnicky97a}, and references
therein). If the northern cloud were pristine infalling material, it
would be highly unlikely to share the same gas-phase metal abundance
as the southern \HI\ component.  If the northern component became
kinematically distinct as a result of SF ongoing over the past 500
Myr, then the identical abundances are to be expected.

\subsection{Regions Without Coherent \HI\ Structure}
\label{S5.2}

While the evidence in \S\ref{S5.1.1} argues in favor of a stellar
feedback origin for the \HI\ holes, we are left with an equally
challenging question: why are coherent \HI\ structures created in some
regions, but not in others with essentially identical physical
characteristics?  Stated equivalently, why do we not observe
\HI\ holes associated with some of the more active regions studied
here?  Consider, in this context, a comparison of Hole 2 and Control
13.  They share similar average \HI\ column densities, UV surface
brightnesses, and stellar densities.  Their CMDs are similar, with
strong BHeB and RHeB loci; their SFHs share the same qualitative
shapes and normalized SFR values.  They both have elevated SFRs in the
100-25 Myr interval.  While the recent SF episode is stronger in Hole
2 than in Control 13, Hole 2 covers a slightly larger physical area.
Control 13 does not create a coherent \HI\ structure, but Hole 2 does.

Given the global burst of SF that DDO\,165 has undergone in the last 1
Gyr (see {McQuinn \etal\ 2010}\nocite{mcquinn10}), and the energies
available from stellar evolution processes in the regions studied
here, it may in fact seem surprising that so few coherent
\HI\ structures are detected in DDO\,165. To explain this, we propose
an alternative interpretation, where the ongoing SF (which changes
strength as a function of time, position and thus physical
characteristics) is capable of creating incoherent but bulk motion of
the interstellar material surrounding localized SFR peaks.  This
provides a simple explanation for the prevalence of the localized high
velocity \HI\ gas motions that permeate the southern \HI\ component
(see Paper I).  When SF is spatially concentrated and temporally
extended, it is capable of creating macroscopic \HI\ holes and shells.
When SF permeates large regions with variations in intensity (e.g.,
throughout the southern \HI\ component, where relatively few
\HI\ holes are found compared to the number of control and cluster
regions), the output energies serve to create incoherent but high
velocity motions of the \HI\ gas when seen at the physical resolution
of these data.  This suggests that, in addition to the classical
\HI\ holes and shells, localized high velocity gas motions can be
interpreted as a signature of stellar feedback.  This is in agreement
with the modeling results of \citet{joung06} and \citet{joung09},
where the local velocity dispersion increases as a result of elevated
SFRs.  Further simulations of the interaction of spatially and
extended SF on the surrounding ISM, with various assumptions on the
efficiency and mechanisms of coupling between the energy and the gas,
would be very valuable.

We conclude that regions with significant SF activity that lack
coherent \HI\ structures are a natural product of temporally and
spatially distributed SF.  This is based on multiple observed and
theoretical properties of the ISM.  First and most obviously, coherent
\HI\ motions smaller than our resolution limit ($\sim$160 pc) cannot
be discerned in these data.  Second, spatially extended SF episodes
can have both creative and destructive effects on coherent
\HI\ structures.  An examination of the movies discussed in
\S~\ref{S4.1} shows numerous examples of nearby regions of elevated
normalized SF rates; coherent motions from one may encounter opposing
motions from another.  Third, the porosity of the ISM is challenging
to measure at the present time \citep[e.g.,][]{heckman01} and is
unknown prior to the onset of SF in any given region. Feedback energy
deposited into a porous region will more efficiently escape than in a
uniform surrounding medium \citep[e.g.,][]{kunth98}.  Finally, the
position of the SF region with respect to the galactic midplane is an
important parameter in the evolution of expanding structures from SF
sites \citep[e.g.,][]{maclow99}.  Taken as a whole, these arguments
favor a complex scenario for the deposition of feedback energy into
the surrounding ISM; the efficiency of this process will depend on
many factors (e.g., the intensity, duration and distribution of
massive star formation, the metal content of the parent cloud, the
local and total gravitational potential depth and shape, the local
neutral and molecular gas density, the ISM porosity, position with
respect to galactic midplane, etc.), leading naturally to both
coherent \HI\ holes and to incoherent motion of the interstellar
material.

\section{Conclusions}
\label{S6}

We have presented a comparison of the recent SFH in DDO\,165 with the
complex neutral gas morphology and kinematics.  Using {\it HST} imaging
we construct CMDs that allow us to isolate the age-sensitive helium
burning stars.  We exploit the fine temporal resolution back to 500
Myr in order to create both global and localized SFHs of selected
regions within DDO\,165.  By comparison with synthesis models and
stellar evolution isochrones, these SFHs can be used to derive the
amount of energy unleashed by evolving stars into their immediate
surroundings as a function of time.

The global SFH of DDO\,165 reveals interesting evolutionary
characteristics.  Compared to the lifetime average SFR
(1.3\,$\times$\,10$^{-2}$ \msun\,yr$^{-1}$), the average SFR over the
last 500 Myr is elevated by a factor of $\sim$3.  This verifies the
remarkably long burst of SF that DDO\,165 has undergone ($>$1 Gyr; see
{McQuinn \etal\ 2010}\nocite{mcquinn10}).  Two intervals of heightened
SFR occur between look back times of 400-300 Myr and 100-25 Myr.  The
SFR drops precipitously in the most recent 25 Myr, consistent with the
system's ``post-starburst'' classification \citep{lee09}.  It is
important to understand how the significant feedback from this
extended starburst episode has impacted the ISM of DDO\,165.

Of specific interest are the \HI\ holes identified in Paper I; our
spatially resolved {\it HST} data allow us to examine the stellar
populations within these structures and to quantify their SFHs and
feedback energy budgets.  In addition to these \HI\ holes, we study
the stellar populations in 2 regions containing stellar clusters, one
region associated with the kinematically distinct northern \HI\ cloud
(with associated stellar cluster), and 20 control fields that sample a
wide variety of optical and UV surface brightnesses, stellar
densities, and \HI\ surface densities.  Many of these control fields
sample regions containing high velocity gas features (from Paper
I). We are thus able to quantitatively examine the localized effects
of massive star evolution.

The most dramatic ISM feature in DDO\,165 is a giant \HI\ hole that
encompasses $\sim$20\% of all of the measured stars within the galaxy.
This structure spans $\sim$2.2\,$\times$\,1.1 kpc and requires
$\sim$10$^{53}$ erg of energy in order to create it.  Our {\it HST}
data reveal that the stellar populations inside of this structure are
energetically capable of creating this structure; in less than
20 Myr there is sufficient feedback energy to meet this requirement.
Extending to a lookback time of 100 Myr, there is nearly 100 times
more feedback energy produced by evolving stars than are required to
create this structure.  Applying an identical analysis to the stars
within each of the other holes leads to the same conclusion: feedback
energies exceed the requisite creation energies, often by large
amounts.

The other localized regions studied within DDO\,165 (i.e., the
``Cluster'', ``Northern'', and ``Control'' fields) have comparable
cumulative feedback energies as the \HI\ holes.  No coherent
\HI\ holes or shells are observed in these regions; this complicates
the stellar feedback interpretation for the creation of HI holes.  We
thus propose an alternative interpretation, where ongoing SF is
capable of creating incoherent but bulk motion of the interstellar
material surrounding localized SFR peaks.  This provides a simple
explanation for the prevalence of the localized high velocity \HI\ gas
motions that permeate the southern \HI\ component (see Paper I).  When
SF is spatially concentrated and temporally extended, it is capable of
creating macroscopic \HI\ holes and shells.  When SF permeates large
regions with variations in intensity, the output energies serve to
create incoherent but high velocity motions of the \HI\ gas.  We
conclude that regions with significant SF activity that lack coherent
\HI\ structures are a natural product of temporally and spatially
distributed SF.

Perhaps as remarkable as the giant \HI\ hole and the high velocity gas
features is the kinematically distinct nature of the northern
\HI\ complex.  As demonstrated in Paper I, this region is static along
the line of sight with respect to the main body of DDO\,165.  The
northern complex contains $\sim$15\% of the total \HI\ mass in the
system; column densities exceed 10$^{21}$ cm$^{-2}$ and young stars
are associated with this gas cloud.  Using \HI\ data alone, both an
infall and an outflow origin for this kinematic discontinuity remain
viable.  The energies available from our SFH analysis, as well as
the chemical homogeneity found by \citet{croxall09}, favor a scenario
where the northern complex has a feedback-triggered outflow origin,
with in situ SF proceeding in the ejected gas.  Stated equivalently,
the $\sim$10$^{56}$ erg of feedback energy produced over the last 500
Myr by the ongoing starburst is energetically capable of creating the
giant \HI\ hole, ejecting the northern gas cloud away from the center
of the system, and inducing incoherent but high velocity motion of a
large fraction of the neutral gas.

The giant \HI\ hole in DDO\,165 is similar to the ones found in the
M\,81 group dwarfs M81\,dwA and Ho\,I, studied in detail by
\citet{warren11}.  As that work and the present results indicate,
feedback from recent SF is energetically capable of producing each of
these kpc-size structures. DDO\,165 and Ho\,I have neutral gas masses
in the 10$^8$ \msun\ range; a characteristic cumulative feedback
energy of 10$^{55}$ deposited on 100 Myr timescales is capable of
creating kpc-size \HI\ holes.  The neutral gas mass and cumulative
feedback energy in M81\,dwA are each an order of magnitude lower over
the same characteristic time interval.  While our SFH analysis allows
us to conclude that SF can create such giant \HI\ structures, we are
left puzzled as to why it does not do so in other M\,81 group dwarf
galaxies with similar normalized SFRs and neutral gas masses.  The
creation of giant \HI\ structures appears to be efficient in galaxies
less massive than $\sim$10$^9$ \msun\ that host temporally extended SF
events that include a significant fraction (perhaps 30\% or more) of
all stars in the system.

DDO\,165 is an extreme example of how feedback drives the evolution of
dwarf galaxies.  The Gyr-long burst clearly indicates that low-mass
galaxies are capable of vigorous SF over timescales much longer than
the canonical instantaneous burst.  The resulting energy outputs
from such long SF episodes are sufficient to induce significant bulk
motions in the surrounding ISM.  The signatures of feedback in
DDO\,165 are diverse, including coherent and incoherent gas motions.
DDO\,165 serves as a poignant reminder that a remarkably active past
can be unlocked by probing the SFHs of nearby galaxies.

\acknowledgements

We thank an anonymous referee for an especially detailed, thorough and
insightful report that helped to significantly strengthen this
manuscript.  Support for this work was provided by NASA through grant
GO-10605 from the Space Telescope Science Institute, which is operated
by AURA, Inc., under NASA contract NAS5-26555.  J.M.C. and
H.P.M. thank Macalester College for research support.  D.R.W. and
S.R.W. are grateful for support from Penrose Fellowships.  This
research has made use of the NASA/IPAC Extragalactic Database (NED)
which is operated by the Jet Propulsion Laboratory, California
Institute of Technology, under contract with the National Aeronautics
and Space Administration, and NASA's Astrophysics Data System.

\bibliographystyle{apj}                                                 


\clearpage
\begin{deluxetable}{lccccccc}
\tabletypesize{\scriptsize}
\tablecaption{\HI\ Holes in DDO\,165 Within the HST Field of View} 
\tablewidth{0pt}  
\tablehead{ 
\colhead{Number}    &\colhead{$\alpha$}                 &\colhead{$\delta$}                 &\colhead{Diameter}              &\colhead{Geometric}                           &\colhead{Galactocentric}                  &\colhead{V$_{\rm exp}$}                        &\colhead{Kinematic Age}\\
\colhead{}          &\colhead{(J2000)\tablenotemark{a}} &\colhead{(J2000)\tablenotemark{a}} &\colhead{(pc)\tablenotemark{b}} &\colhead{Mean Radius\tablenotemark{c}\, (pc)} &\colhead{Radius\tablenotemark{d}\, (kpc)} &\colhead{(km\,s$^{-1}$)\tablenotemark{e}} &(Myr)\tablenotemark{f}}
\startdata      
\vspace{0.0 cm}      
1 &13:06:17.1 &67:43:04.4 &975\,$\times$\,325, 40\degr    &281  &1.16 &-  &-    \\
2 &13:06:22.1 &67:41:46.9 &520                            &260  &1.05 &11 &23.0 \\
4 &13:06:27.7 &67:41:04.5 &390                            &195  &1.96 &-  &-    \\
5 &13:06:30.1 &67:41:26.5 &737                            &369  &1.59 &7  &51.2 \\
6 &13:06:30.6 &67:42:31.0 &2210\,$\times$\,1084, 130\degr &774  &0.73 &-  &-    \\
7 &13:06:38.6 &67:41:33.7 &455                            &228  &2.14 &9  &24.6 \\
\enddata     
\label{t1}
\tablenotetext{a}{The right ascension and declination of the center of
the holes. The error of these values is \pom\ 3\arcsec. Units of right
ascension are hours, minutes, and seconds, and units of declination
are degrees, arcminutes and arcseconds.}  
\tablenotetext{b}{The diameter of the hole in pc. If the hole is
elliptical then the major and minor axis diameters, and the major axis 
position angle (east of north), are also listed.  The hole diameters 
listed have an error of \pom\ 100 pc}
\tablenotetext{c}{The radius of a circle with area equivalent to that of the ellipse; this radius is used in energy calculations.}
\tablenotetext{d}{Radial distance from the assumed center of the \HI\ distribution ($\alpha$,$\delta$ = 13:06:24.7, 67:42:33.2).}
\tablenotetext{e}{Expansion velocity of the hole, when it can be
measured, with an error of 2 \kms.}
\tablenotetext{f}{The calculated ages for the holes are upper limits
as we assumed the expansion velocity to be constant throughout its evolution.}
\end{deluxetable}   

\clearpage
\begin{deluxetable}{lcccc}  
\tabletypesize{\small} 
\tablecaption{Properties of Selected Regions Within DDO\,165} 
\tablewidth{0pt}  
\tablehead{ 
\colhead{Identification} &\colhead{$\alpha$} &\colhead{$\delta$} &\colhead{Diameter} &\colhead{Number of Stars}\\
\colhead{}               &\colhead{(J2000)}  &\colhead{(J2000)}  &\colhead{(arcsec)} &\colhead{in CMD}}
\startdata      
\vspace{0.0 cm}      
\HI\ Hole 1                               &13:06:17.1 &67:43:04.4 &45$\times$15, 40\degr   &2888\\
\HI\ Hole 2                               &13:06:22.1 &67:41:46.9 &24                      &2690\\
\HI\ Hole 4                               &13:06:27.7 &67:41:04.5 &18                      &310\\
\HI\ Hole 5                               &13:06:30.1 &67:41:26.5 &34                      &2126\\
\HI\ Hole 6                               &13:06:30.6 &67:42:31.0 &102$\times$50, 310\degr &28439\\
\HI\ Hole 7                               &13:06:38.6 &67:41:33.7 &21                      &701\\
Cluster\,1 (CL\,1)               &13:06:14.6 &67:42:23.5 &20                      &2307\\
Cluster\,2 (CL\,2)                &13:06:17.8 &67:42:13.5 &20                      &2321\\
Northern\,1 (N1)                  &13:06:24.4 &67:43:41.2 &12                      &261\\
Northern\,2 (N2)                  &13:06:24.4 &67:43:41.2 &24                      &957\\
Control Region\,1    (C1)           &13:06:11.1 &67:42:32.9 &20                      &1901\\
Control Region\,2    (C2)           &13:06:12.1 &67:43:03.4 &20                      &1274\\
Control Region\,3    (C3)           &13:06:12.4 &67:42:02.8 &20                      &1631\\
Control Region\,4    (C4)           &13:06:16.6 &67:43:38.0 &20                      &615\\
Control Region\,5    (C5)           &13:06:17.0 &67:41:48.4 &20                      &1573\\
Control Region\,6    (C6)           &13:06:17.1 &67:41:23.7 &20                      &2224\\
Control Region\,7    (C7)           &13:06:17.4 &67:42:37.4 &20                      &681\\
Control Region\,8    (C8)           &13:06:21.0 &67:42:47.9 &20                      &2056\\
Control Region\,9    (C9)           &13:06:21.6 &67:42:24.5 &20                      &2566\\
Control Region\,10   (C10)          &13:06:23.1 &67:43:17.4 &20                      &1366\\
Control Region\,11   (C11)          &13:06:24.5 &67:41:20.8 &20                      &730\\
Control Region\,12   (C12)          &13:06:24.6 &67:42:08.5 &20                      &2410\\
Control Region\,13   (C13)          &13:06:27.9 &67:41:54.1 &20                      &2368\\
Control Region\,14   (C14)          &13:06:29.8 &67:43:29.1 &20                      &684\\
Control Region\,15   (C15)          &13:06:34.3 &67:43:16.2 &20                      &1008\\
Control Region\,16   (C16)          &13:06:35.1 &67:41:22.5 &20                      &534\\
Control Region\,17   (C17)          &13:06:37.4 &67:42:52.4 &20                      &1539\\
Control Region\,18   (C18)          &13:06:40.2 &67:41:58.4 &20                      &1300\\
Control Region\,19   (C19)          &13:06:41.0 &67:42:27.5 &20                      &1354\\
Control Region\,20   (C20)          &13:06:42.1 &67:43:07.8 &20                      &718\\
\enddata  \vspace{-0.5 cm}   
\label{t2}
\end{deluxetable}   

\clearpage
\begin{deluxetable}{lcccc}  
\tabletypesize{\small} 
\tablecaption{Properties of Stellar Clusters Within DDO\,165} 
\tablewidth{0pt}  
\tablehead{ 
\colhead{Identification} &\colhead{$\alpha$} &\colhead{$\delta$} &\colhead{Age}   &\colhead{Mass}    \\
\colhead{}               &\colhead{(J2000)}  &\colhead{(J2000)}  &\colhead{(yr)}  &\colhead{(\msun)}}
\startdata      
\vspace{0.0 cm}      
c1& 13:06:10.2 & 67:41:42.2  &  (6.3\pom3.3)\,$\times$\,10$^8$    &(2.0\pom0.7)\,$\times$\,10$^4$  \\
c2& 13:06:13.9 & 67:42:22.5  &  (1.3\pom0.2)\,$\times$\,10$^9$    &(9.1\pom1.4)\,$\times$\,10$^4$  \\
c3& 13:06:14.6 & 67:41:30.5  &  (2.2\pom1.3)\,$\times$\,10$^9$    &(3.6\pom1.6)\,$\times$\,10$^4$  \\
c4& 13:06:14.8 & 67:43:13.4  &  no fit                            &no fit\\
c5& 13:06:15.4 & 67:42:20.1  &  (2.5\pom0.2)\,$\times$\,10$^8$    &(6.8\pom0.4)\,$\times$\,10$^4$  \\
c6& 13:06:16.3 & 67:42:05.8  &  (3.2\pom1.2)\,$\times$\,10$^8$    &(2.0\pom0.4)\,$\times$\,10$^4$  \\
c7& 13:06:17.9 & 67:42:13.9  &  (1.1\pom0.3)\,$\times$\,10$^8$    &(6.1\pom1.0)\,$\times$\,10$^4$  \\
c8& 13:06:18.1 & 67:41:34.7  &  no fit                            &  no fit    \\
c9& 13:06:20.0 & 67:43:49.4  &  (1.6\pom1.2)\,$\times$\,10$^8$    &(5.2\pom1.9)\,$\times$\,10$^3$  \\
c10& 13:06:22.6 & 67:42:02.3  & (1.0\pom0.1)\,$\times$\,10$^8$    &(2.5\pom0.1)\,$\times$\,10$^4$  \\
c11& 13:06:23.7 & 67:42:37.9  & (5.0\pom0.2)\,$\times$\,10$^8$    &(3.7\pom0.7)\,$\times$\,10$^4$  \\
c12& 13:06:24.0 & 67:42:58.5  & (1.3\pom0.3)\,$\times$\,10$^9$    &(2.8\pom0.6)\,$\times$\,10$^4$  \\
c13& 13:06:25.8 & 67:42:31.2  & (1.1\pom0.5)\,$\times$\,10$^8$    &(1.2\pom0.3)\,$\times$\,10$^4$  \\
c14& 13:06:29.1 & 67:42:27.2  & (5.6\pom0.3)\,$\times$\,10$^9$    &(5.2\pom0.1)\,$\times$\,10$^5$  \\
c15& 13:06:29.3 & 67:41:28.7  & (5.6\pom2.2)\,$\times$\,10$^9$    &(6.7\pom2.2)\,$\times$\,10$^4$  \\
c16& 13:06:29.8 & 67:42:14.8  & (8.9\pom2.5)\,$\times$\,10$^7$    &(7.2\pom0.9)\,$\times$\,10$^3$  \\
c17& 13:06:30.2 & 67:41:52.0  & (1.4\pom0.5)\,$\times$\,10$^8$    &(1.6\pom0.2)\,$\times$\,10$^4$  \\
c18& 13:06:30.8 & 67:42:17.2  & (2.2\pom0.3)\,$\times$\,10$^9$    &(3.6\pom0.4)\,$\times$\,10$^5$  \\
c19& 13:06:30.8 & 67:42:41.2  & (1.3\pom0.4)\,$\times$\,10$^8$    &(1.0\pom0.1)\,$\times$\,10$^4$  \\
c20& 13:06:35.8 & 67:42:07.1  & (3.2\pom1.0)\,$\times$\,10$^8$    &(1.3\pom0.2)\,$\times$\,10$^4$  \\
\enddata  \vspace{-0.5 cm}   
\label{t3}
\end{deluxetable}

\clearpage
\begin{deluxetable}{lccccccccc}  
 \tabletypesize{\footnotesize} 
\tablecaption{Energy Estimates for \HI\ Holes With Measured Expansion Velocities} 
\tablewidth{0pt}  
\tablehead{ 
\colhead{Hole} &\colhead{V$_{\rm exp}$} &\colhead{$n_0$\tablenotemark{a}} &\colhead{Energy\tablenotemark{a}}          &\colhead{$n_0$\tablenotemark{b}} &\colhead{Energy\tablenotemark{b}}          &\colhead{$n_0$\tablenotemark{c}} &\colhead{Energy\tablenotemark{c}}          &\colhead{$n_0$\tablenotemark{d}} &\colhead{Energy\tablenotemark{d}}\\
\colhead{}     &\colhead{(\kms)}        &\colhead{(cm$^{-3}$)}            &\colhead{($10^{51}$erg)} &\colhead{(cm$^{-3}$)}            &\colhead{($10^{51}$erg)} &\colhead{(cm$^{-3}$)}            &\colhead{($10^{51}$erg)} &\colhead{(cm$^{-3}$)}            &\colhead{($10^{51}$erg)}\\
\colhead{(1)} &\colhead{(2)} &\colhead{(3)} &\colhead{(4)} &\colhead{(5)} &\colhead{(6)}  &\colhead{(7)}  &\colhead{(8)} &\colhead{(9)} &\colhead{(10)}}
\startdata      
\vspace{0.0 cm} 
2              &11                      &0.1                             &4.0                      &0.094                            &3.7                       &0.091                            &3.6                       &0.094                            &3.7\\
5              &7                       &0.1                             &6.2                      &0.078                            &4.7                       &0.078                            &4.7                       &0.073                            &4.3\\
7              &9                       &0.1                             &2.0                      &0.051                            &0.92                      &0.053                            &0.96                      &0.049                            &0.90\\
\enddata       
\label{t4}
\tablenotetext{a}{Energy calculation from Paper I assuming $n_0$ = 0.1}
\tablenotetext{b}{From best fit line (red curve in Figure~\ref{figcap18}).}
\tablenotetext{c}{From best fit line, extrapolated assuming a constant volume density in the central regions (blue curve in Figure~\ref{figcap18}).}
\tablenotetext{d}{From best fit Gaussian curve (green curve in Figure~\ref{figcap18}).}
\end{deluxetable}   

\clearpage
\begin{deluxetable}{lccccccccc} 
\tabletypesize{\footnotesize} 
\tablecaption{Energy Estimates for \HI\ Holes With No Measured Expansion Velocities}
\tablewidth{0pt}  
\tablehead{ 
\colhead{Hole} &\colhead{V$_{\rm exp}$} &\colhead{$n_0$\tablenotemark{a}} &\colhead{Energy\tablenotemark{a}}          &\colhead{$n_0$\tablenotemark{b}} &\colhead{Energy\tablenotemark{b}}          &\colhead{$n_0$\tablenotemark{c}} &\colhead{Energy\tablenotemark{c}}          &\colhead{$n_0$\tablenotemark{d}} &\colhead{Energy\tablenotemark{d}}\\
\colhead{}     &\colhead{(\kms)}        &\colhead{(cm$^{-3}$)}            &\colhead{($10^{51}$erg)} &\colhead{(cm$^{-3}$)}            &\colhead{($10^{51}$erg)} &\colhead{(cm$^{-3}$)}            &\colhead{($10^{51}$erg)} &\colhead{(cm$^{-3}$)}            &\colhead{($10^{51}$erg)}\\
\colhead{(1)} &\colhead{(2)} &\colhead{(3)} &\colhead{(4)} &\colhead{(5)} &\colhead{(6)}  &\colhead{(7)}  &\colhead{(8)} &\colhead{(9)} &\colhead{(10)}}
\startdata 
\multicolumn{10}{c}{Assuming V$_{\rm exp}$ = 7 \kms, as in THINGS analysis.}\\ 
\tableline\\
1              &7                       &0.1                              &2.7                       &0.094                            &2.5                       &0.090                            &2.4                        &0.090                            &2.4\\
4              &7                       &0.1                              &0.86                      &0.059                            &0.47                      &0.062                            &0.50                       &0.057                            &0.46\\
6              &7                       &0.1                              &63                        &0.081                            &50                        &0.094                            &59                       &0.11                            &67\\
\tableline\\
\multicolumn{10}{c}{Assuming V$_{\rm exp}$ = $\sigma_{\rm V}$ (velocity dispersion) over the entire hole.}\\ 
\tableline\\
1              &11.2                    &0.1                              &5.2                       &0.094                            &4.8                       &0.090                            &4.6                       &0.090                            &4.6\\
4              &12.0                    &0.1                              &1.8                       &0.059                            &1.0                       &0.062                            &1.1                       &0.057                            &0.97 \\
6              &12.0                    &0.1                              &130                       &0.081                            &110                       &0.094                            &130                       &0.11                            &140\\
\tableline\\ 
\multicolumn{10}{c}{Assuming V$_{\rm exp}$ = $\sigma_{\rm V}$ (velocity dispersion) over the entire galaxy.}\\ 
\tableline\\
1              &8.5                     &0.1                              &3.5                       &0.094                            &3.3                       &0.090                            &3.1                       &0.090                            &3.1\\
4              &8.5                     &0.1                              &1.1                       &0.059                            &0.62                      &0.062                            &0.66                      &0.057                            &0.60\\
6              &8.5                     &0.1                              &83                        &0.081                            &65                        &0.094                            &77                      &0.11                            &87\\
\enddata     
\label{t5}
\tablenotetext{a}{Assuming a constant volume density (n$_{\rm 0}$ = 0.1 cm$^{-3}$) throughout the galaxy; see Paper I.}
\tablenotetext{b}{From best fit line (red curve in Figure~\ref{figcap18}).}
\tablenotetext{c}{From best fit line, extrapolated assuming a constant volume density in the central regions (blue curve in Figure~\ref{figcap18}).}
\tablenotetext{d}{From best fit Gaussian curve (green curve in Figure~\ref{figcap18}).}
\end{deluxetable}  

\clearpage
\begin{deluxetable}{lccccc}
\tabletypesize{\scriptsize}
\tablecaption{Quantifying Feedback in DDO\,165}
\tablewidth{0pt}  
\tablehead{ 
\colhead{Number} &\colhead{Adopted}                    &\colhead{E$_{\rm Hole}$\tablenotemark{b}} &\colhead{{E$_{\rm SF}$}\tablenotemark{c}}   &\colhead{Feedback}                     &\colhead{t$_{\rm eq}$\tablenotemark{e}}\\
\colhead{}       &\colhead{Age\tablenotemark{a} \,\,(Myr)} &\colhead{($10^{51}$erg)}                 &\colhead{($10^{51}$erg)}                   &\colhead{Efficiency\tablenotemark{d}}  &\colhead{(Myr)}}
\startdata      
\vspace{0.0 cm}    
1                &32                                   &3.1                                       &8.4                                         &37\%                                   &$<$30 Myr\\
2                &23                                   &3.7                                       &28                                          &13\%                                   &$<$15 Myr\\
4                &22                                   &0.06                                      &N/A                                         &N/A                                    &$\sim$330 Myr\\
5                &52                                   &4.3                                       &76                                          &5.7\%                                  &$<$25 Myr\\
6                &89                                   &87.4                                      &6170                                        &1.4\%                                  &$<$20 Myr\\
7                &25                                   &0.90                                      &N/A                                         &N/A                                    &$\sim$130 Myr\\
\enddata     
\label{t6}
\tablenotetext{a}{Calculated using the geometric radius of each hole, the measured expansion velocities for Holes 2, 5, and 7, and the 
total galaxy velocity dispersion for Holes 1, 4 and 6.}
\tablenotetext{b}{The values use the Gaussian fit for volume densities of all holes; the expansion velocities are either direct measurements 
(Holes  2, 5, 7) or are adopted to be equivalent to the global $\sigma_{\rm V}$ value (8.5 \kms) for non-expanding holes (Holes 1, 4,  6).}
\tablenotetext{c}{The total energy input by stellar evolution over the kinematic age of the \HI\ hole, computed using STARBURST99.}
\tablenotetext{d}{The ratio E$_{\rm Hole}$/E$_{\rm SF}$.}
\tablenotetext{e}{The time when E$_{\rm SF}$ $=$ E$_{\rm Hole}$.}
\end{deluxetable}   

\clearpage
\begin{deluxetable}{lccc}  
\tabletypesize{\small} 
\tablecaption{Feedback Energies in Selected Regions Within DDO\,165} 
\tablewidth{0pt}  
\tablehead{ 
\colhead{Identification}          &\colhead{E$_{\rm SF}$ 25 Myr}               &\colhead{E$_{\rm SF}$ 50 Myr}                  &\colhead{E$_{\rm SF}$ 100 Myr}\\
\colhead{}                        &\colhead{(10$^{51 }$erg)\tablenotemark{a}}  &\colhead{(10$^{51 }$erg)\tablenotemark{b}}    &\colhead{(10$^{51 }$erg)\tablenotemark{c}}}
\startdata      
\vspace{0.0 cm}      
\HI\ Hole 1                       &   2.5 &   24     &  110  \\
\HI\ Hole 2                       &  28   &  270     & 1400  \\
\HI\ Hole 4                       &   0.0 &    0.0   &    0.0\\
\HI\ Hole 5                       &   6.8 &   60     &  230  \\
\HI\ Hole 6                       & 210   & 1700     & 7500  \\
\HI\ Hole 7                       &   0.0 &    0.0   &    0.4\\
Cluster\,1 (CL\,1)                &  33   &  290     &  980  \\
Cluster\,2 (CL\,2)                & 280   & 1100     & 2400  \\
Northern\,1 (N1)                  &  15   &   75     &  180  \\
Northern\,2 (N2)                  &  34   &  130     &  270  \\
Control Region\,1  (C1)           &   6.5 &   58     &  240  \\
Control Region\,2  (C2)           &   1.5 &   13     &   47  \\
Control Region\,3  (C3)           &   4.9 &   44     &  170  \\
Control Region\,4  (C4)           &   0.9 &    8.1   &   29  \\
Control Region\,5  (C5)           &  21   &   51     &  160  \\
Control Region\,6  (C6)           &   0.4 &    3.3   &    9.6\\
Control Region\,7  (C7)           &  15   &  130     &  470  \\
Control Region\,8  (C8)           &   3.0 &   29     &  160  \\
Control Region\,9  (C9)           &  17   &  150     &  500  \\
Control Region\,10 (C10)          &  42   &  110     &  190  \\
Control Region\,11 (C11)          &   0.0 &    0.0   &    0.0\\
Control Region\,12 (C12)          &  70   &  380     & 1300  \\
Control Region\,13 (C13)          &  20   &  210     & 1100  \\
Control Region\,14 (C14)          &   2.0 &   17     &   48  \\
Control Region\,15 (C15)          &   0.0 &    0.4   &    9.7\\
Control Region\,16 (C16)          &   0.0 &    0.1   &    2.5\\
Control Region\,17 (C17)          &   7.7 &   67     &  250  \\
Control Region\,18 (C18)          &  49   &  190     &  510  \\
Control Region\,19 (C19)          &   7.5 &   69     &  320  \\
Control Region\,20 (C20)          &   0.0 &    0.2   &    5.0\\
Total Galaxy                      & 1900  &11000     &35000\\
\enddata  
\label{t7}
\tablenotetext{a}{Energy from SF over the last 25 Myr.}
\tablenotetext{b}{Energy from SF over the last 50 Myr.}
\tablenotetext{c}{Energy from SF over the last 100 Myr.}
\end{deluxetable}   

\clearpage

\clearpage

\clearpage

\clearpage

\clearpage

\clearpage

\clearpage

\clearpage

\clearpage

\clearpage

\clearpage

\clearpage

\clearpage

\clearpage

\clearpage

\clearpage

\clearpage

\clearpage

\clearpage

\clearpage

\clearpage

\clearpage

\end{document}